\documentclass[epj,nopacs]{svjour}
\usepackage{graphics}
\usepackage{adjustbox}
\usepackage{siunitx}
\usepackage{subfig}
\usepackage{hyperref}
\usepackage{lineno}
\usepackage{textgreek}
\usepackage{multicol}
\usepackage{enumitem}
\usepackage{subfiles}
\usepackage{diagbox}
\usepackage{bm}
\usepackage{cite}
\usepackage{amsmath}
\usepackage{amssymb}
\usepackage{soul}
\usepackage{tabularx}
\usepackage{float}

\begin{document}
\title{Energy reconstruction for large liquid scintillator detectors with machine learning techniques: aggregated features approach}

\author{Arsenii Gavrikov\inst{1, 2,}\thanks{gavrikov@jinr.ru (corresponding author). Now at Dipartimento di Fisica e Astronomia dell’Università di Padova and INFN Sezione di Padova, Padova, Italy} \and Yury Malyshkin\inst{2,}\thanks{yum@jinr.ru} \and Fedor Ratnikov\inst{1,}\thanks{fratnikov@hse.ru}
}                     
%
%
\institute{HSE University, Moscow, Russia \and Joint Institute for Nuclear Research, Dubna, Russia}

\date{Received: date / Revised version: date}
%
\abstract{
Large-scale detectors consisting of a liquid scintillator target surrounded by an array of photo-multiplier tubes (PMTs) are widely used in the modern neutrino experiments: Borexino, KamLAND, Daya Bay, Double Chooz, RENO, and the upcoming JUNO with its satellite detector TAO. Such apparatuses are able to measure neutrino energy which can be derived from the amount of light and its spatial and temporal distribution over PMT channels. However, achieving a fine energy resolution in large-scale detectors is challenging.
In this work, we present machine learning methods for energy reconstruction in the JUNO detector, the most advanced of its type. 
We focus on positron events in the energy range of 0--10~MeV which corresponds to the main signal in JUNO --- neutrinos originated from nuclear reactor cores and detected via the inverse beta decay channel. We consider the following models: Boosted Decision Trees and Fully Connected Deep Neural Network, trained on aggregated features, calculated using the information collected by PMTs. We describe the details of our feature engineering procedure and show that machine learning models can provide the energy resolution $\sigma = 3\%$ at 1~MeV using subsets of engineered features. The dataset for model training and testing is generated by the Monte Carlo method with the official JUNO software.
}

\maketitle

\section{Introduction}
\label{sec:intro}

One of neutrino detection methods consists in observation of optical photons emitted in the target volume by all the particles produced as the result of the neutrino interaction.
Although charged particles can produce Cherenkov light even in water, a certain class of detectors, e.g. Borexino~\cite{Borexino:2002},
KamLAND~\cite{KamLAND:2002uet}, Daya Bay~\cite{DayaBay:2012fng}, Double Chooz~\cite{DoubleChooz:2011ymz}, and RENO~\cite{RENO:2012mkc}, used liquid scintillators to increase the light yield. Modern compositions of liquid scintillators provide $O(10^4)$ photons per MeV of deposited energy, a significant part of which can be observed with photomultiplier tubes (PMTs) placed around the target. Since the amount of emitted photons is defined by the deposited energy, it can be easily estimated. Using extra information from the temporal distribution and spatial pattern of fired PMTs allows improving the energy measurement accuracy. However, future detectors with a large number of channels (PMTs) will make the task challenging. We consider novel approaches based on machine learning for the detector of the Jiangmen Underground Neutrino Observatory (JUNO)~\cite{JUNO:2015zny, JUNO:2021vlw}, the largest of its type, currently under construction. These approaches can also be applied to other similar detectors.

JUNO is a multipurpose neutrino observatory located in South China. The primary aims of the JUNO experiment are to determine the neutrino mass ordering and to precisely measure neutrino oscillation parameters $\sin^2{\theta_{12}}$, $\Delta m_{21}^2$, $\Delta m^{2}_{31}$. The main source of neutrinos in JUNO will be the Yangjiang and Taishan nuclear power plants located about 52.5~kilometers away from the detector. Concurrently, JUNO will be able to explore neutrinos from supernovae, atmospheric and solar neutrinos, geoneutrinos, as well as some rare processes, like the proton decay.

\begin{figure*}[!htb]
	\centering
	\includegraphics[width=0.85\textwidth]{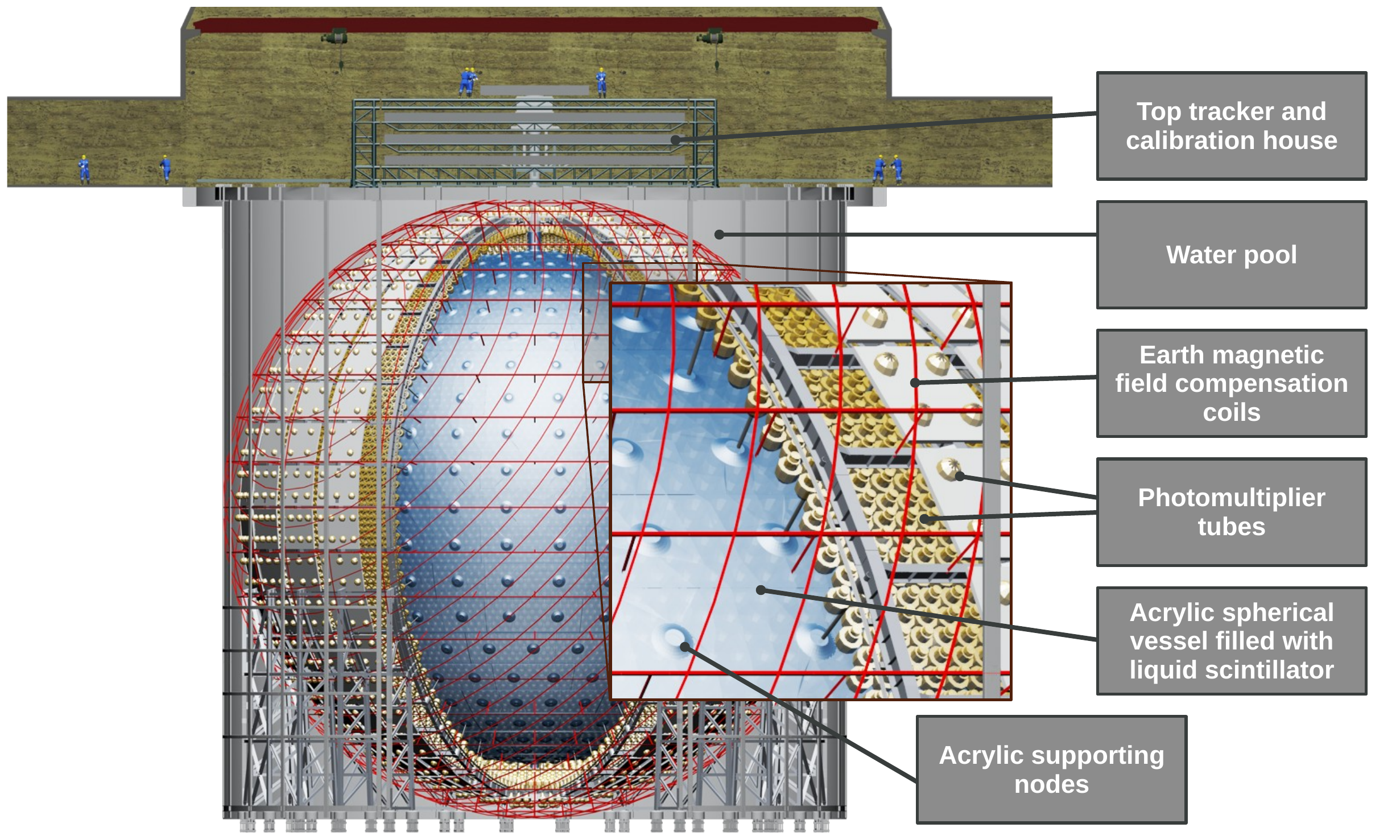}
	\caption{Schematic view of JUNO detector and other main components}
	\label{fig:detector}
\end{figure*}
The key requirement for JUNO is to provide the energy resolution $\sigma_{E}/E = 3\% / \sqrt{E(\text{MeV})}$. The detector construction illustrated in Figure~\ref{fig:detector} is optimized to meet this requirement. The JUNO detector consists of the Central Detector (CD), a water Cherenkov detector, and the Top Tracker. The CD is an acrylic sphere 35.4~meters in diameter filled with 20~kt of liquid scintillator. The CD is held by a stainless steel construction immersed in a water pool. It is equipped with a large number of PMTs of two types: \num{17612} large 20-inch tubes and \num{25600} small 3-inch tubes.
The former provide 75.2\% the sphere coverage, and the latter add extra 2.7\% of coverage~\cite{JUNO:2021vlw}.
The water pool is also equipped with \num{2400} 20-inch PMTs to detect Cherenkov light from muons. The Top Tracker is disposed at the top of the detector and used to detect muon tracks. The Calibration House integrates calibration systems.

JUNO will detect electron antineutrinos via the inverse beta decay (IBD) channel: $\overline \nu_{e} + p \rightarrow e^{+} + n$.
The positron deposits its energy and annihilates into two 0.511~MeV gammas forming a so-called prompt signal. The energy is deposited shortly after the interaction and is the sum of the positron kinetic energy and the annihilation energy of two 0.511 MeV gammas: $E_{\text{dep}} = E_{\text{kin}} + 1.022$~MeV.
The neutron is captured in liquid scintillator by hydrogen or carbon nuclei after approximately 200~\textmu s producing 2.22 MeV (99\% cases) or 4.95 MeV (1\% cases) deexcitation gammas for hydrogen and carbon, respectively. This is called a delayed signal.
The time coincidence of prompt and delay signals makes it possible to separate IBD signals from backgrounds\footnote{However, after application of a set of selection cuts, the remaining fraction of background events is indistinguishable from the signal. For example, in case of the reactor neutrino analysis it amounts to about 17.8\% of signal events~\cite{JUNO:osc_sens2022}.}.
The information collected by PMTs will be used for energy and vertex reconstruction of neutrino interactions.

Machine Learning (ML) has experienced an extraordinary rise in recent years.
High-energy physics (HEP), and neutrino physics in particular, have also proven to be remarkable domains for ML applications, especially supervised learning, due to the availability of a large amount of labeled data produced by simulation.
There are many examples of using ML approaches in HEP: in neutrino experiments, in collider experiments, etc.~\cite{Bourilkov:2019yoi, Schwartz:2021ftp, Guest:2018yhq}.
We also recommend the Living Review of ML Techniques for HEP which tries to include all relevant papers~\cite{HEPML:2021}.

The remaining part of the paper is organized as follows. Section~\ref{sec:problem} states the problem. In Section~\ref{sec:datadesc}, we introduce the data used for this analysis. In Section~\ref{sec:ml}, we present feature engineering and ML approaches used for this study and their application to solve the problem.
The performance of the presented ML models is discussed in Section~\ref{sec:results}.
We summarize the study in Section~\ref{sec:conclusions}.

\section{Problem Statement}
\label{sec:problem}

In this work, we continue studying ML techniques for energy reconstruction in the energy range of 0--10~MeV, covering the region of interest for IBD events from reactor electron antineutrinos (previous work published in~\cite{Qian:2021vnh, Gavrikov:2021ktt}). In general, the deposited energy of an event can be reconstructed from individual PMT signals (charge and time) or using aggregated information from the whole array of PMTs --- ``aggregated features''. In~\cite{Qian:2021vnh}, we presented different approaches based on the information obtained PMT-wise as well as on several basic aggregated features. In the subsequent paper~\cite{Gavrikov:2021ktt}, we, using an optimal subset from a large set of newly engineered aggregated features, demonstrated that this approach can achieve the same performance as the one, based on the PMT-wise gained information. On the other hand, vertex reconstruction requires granular information both with traditional and ML algorithms, see~\cite{Li:2021oos, Li:2022tvg}. The actual research aims to further investigate the potential of the aggregated feature approach and to study two models: Boosted Decision Trees and Fully Connected Deep Neural Network for energy reconstruction in JUNO.

To evaluate the performance of the models, we use two metrics: resolution and bias. They are defined by a Gaussian fit of the distribution of difference between predicted and true energy deposition $(E_{\rm pred} - E_{\rm dep})$ (see Section~\ref{sec:results}). The resolution is deﬁned as $\sigma/E_{\rm dep}$ and the bias as $\mu/E_{\rm dep}$, where $\sigma$ and $\mu$ are the standard deviation and the mean of the Gaussian fit, respectively.

To resolve the primary goal of the JUNO experiment, the determination of the neutrino mass ordering at the level of 3 standard deviations, the energy resolution must be better that or equal to $\sigma = 3\%$ at 1~MeV, and the uncertainty of energy nonlinearity should be less than 1\%~\cite{JUNO:2015zny}. 
Following~\cite{JUNO:2015zny}, we use a simple model to describe the energy resolution $\sigma/E_{\rm dep}$ as a function of energy: 
\begin{align}
\label{eq:appr_func}
\frac{\sigma}{E_{\rm dep}} = \sqrt{\left(\frac{a}{\sqrt{E_{\rm dep}}} \right)^2 + b^2 + \left(\frac{c}{E_{\rm dep}} \right)^2},
\end{align}
with parameters $a, b$, and $c$ interpreted as follows: $a$ is mainly driven by statistics of the true accumulated charge on PMTs, $b$ is related to spatial nonuniformity, and $c$ is associated with the charge from dark noise (introduced in Section~\ref{sec:datadesc}). We extract values of $a$, $b$ and $c$ by fitting the energy resolution curve obtained with our energy reconstruction models (see Section~\ref{sec:results}).

It is convenient to have only one parameter to estimate the entire resolution curve.
The JUNO requirements for the appropriate determination of the neutrino mass ordering could be translated into the following requirement on the effective resolution $\tilde{a}$ as~\cite{JUNO:2020xtj}:

\begin{align}
\label{eq:atilde}
\tilde{a} \equiv \sqrt{(a)^2 + (1.6 \times b)^2 + \left(\frac{c}{1.6}\right)^2} \leq 3\%,
\end{align}
reflecting that the effect of the $b$ term is 1.6 larger and the effect of the $c$ term is 1.6 smaller compared to the effect of the $a$ term.

\section{Data Description}
\label{sec:datadesc}

The analyzed data are generated by the full detector Monte Carlo method using the official JUNO software~\cite{Huang:2017dkh, Lin:2017usg, Lin:2017yxy}.
The detector simulation software is based on the Geant4 framework~\cite{GEANT4:2002zbu, Allison:2016lfl}, with the geometry defined according to the latest design~\cite{Li:2018fny}, and is implemented as a standalone application. 

The simulation begins with the injection of a positron with the kinetic energy in the range from 0 to 10 MeV.
As the neutron, produced in a neutrino interaction, is going to be used only for the offline event selection, it was not simulated for this study.
The positron interacts with liquid scintillator losing its energy.
After it stops (or in-flight), the positron annihilates with an electron in the medium, producing a pair of 511~keV gammas. These two gammas are usually stopped by the target, producing secondary Compton electrons. Electromagnetic energy losses of charged particles (the primary positron and secondary electrons in our case) are accompanied by optical light emission. 
In general, about 90\% of light is produced from the scintillation of the detector target medium, and the other ${\sim}$10\% is produced from Cherenkov radiation.

The produced photons are transported through the materials until they are absorbed, leave the detector, or hit PMTs. We do not take into account any liquid flows or inhomogeneities: despite it may make an impact on the detector performance, it is hard to foresee a realistic scenario at the stage of the experiment preparation. We leave it for future studies when the detector will be constructed and we will have information on the liquid flows, temperature variations, etc. Roughly $\sim$30\% of photons hitting the PMT photocathode lead to the release of photoelectrons which initiate an electric pulse. On average, all PMTs collect about 1500 photoelectrons per 1~MeV of the deposited energy at the center of the detector. PMTs also produce pulses spontaneously, the so-called dark current which constitutes the noise.
JUNO will be equipped with two types of 20-inch PMTs: 25\% produced by Hamamatsu and 75\% by NNVT. The measured dark current rate for the former is 19.3~kHz; the latter has the dark rate of 49.3~kHz~\cite{Abusleme:2022cxy}. Different processes during photo-electron collection and further current amplification, as well as intrinsic effects of electronics, are also simulated. The introduced spread of hit times is 1.3~ns and 7.0~ns for Hamamatsu and NNVT, respectively~\cite{Abusleme:2022cxy}. Then the charge and time information is extracted from PMT pulse shapes and serves as an input for reconstruction algorithms. Then, using interaction kinematics, the antineutrino energy can be calculated assuming the following relation: $E_{\tilde{\nu}_{e}} \approx E_{\rm dep} + 0.8$~MeV.

However, if an event happens near the edge of the detector, less than a meter away, one or both photons can escape the CD without contributing to the light yield. The positron energy will be underestimated in such cases. 
There is also a strong radioactive background at the detector edge which also compromises the energy reconstruction in that area. To avoid these effects, we apply the $R <$ \num{17.2}~m volume cut, i.e.\ discard those events in the outermost \num{0.5}~m layer of the detector.

\begin{figure*}[!htb]
	\centering
	\includegraphics[width=1\textwidth]{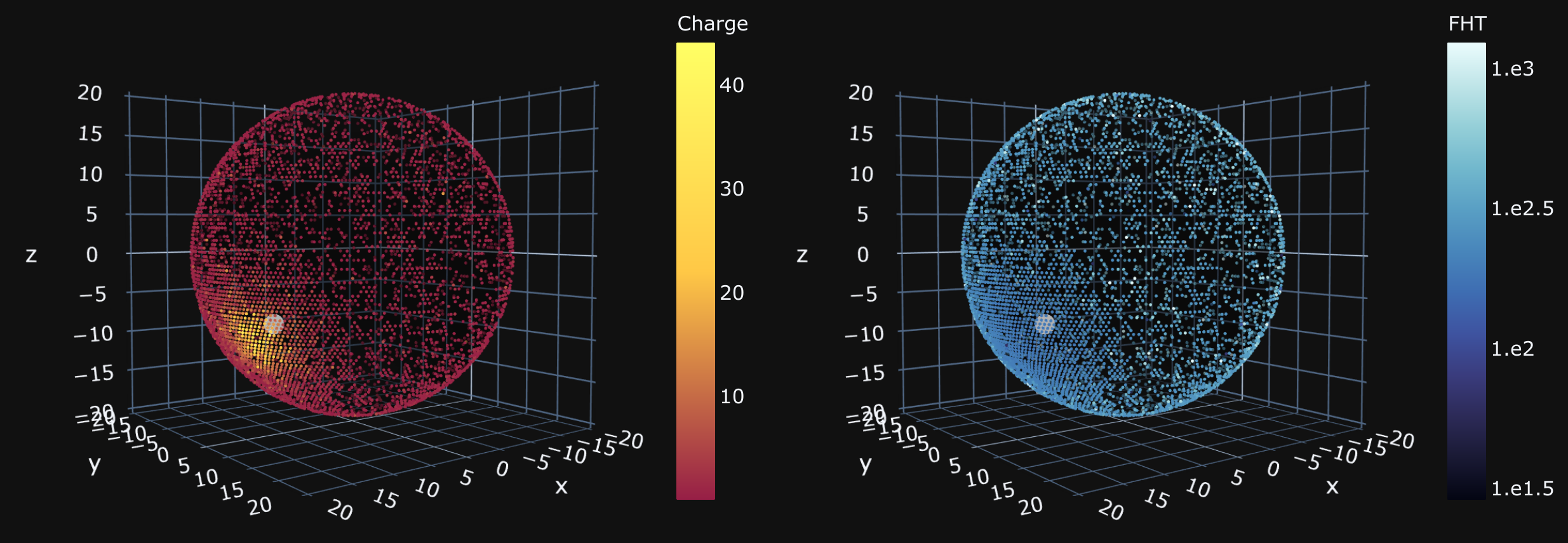}\\
	\caption{Example of a positron event with the deposited energy of 6.165 MeV as seen by 20-inch PMTs. Only fired PMTs are shown. On the left, the color represents the accumulated charge: yellow points show the channels with more hits, red points show the channels with fewer hits. On the right side, the color indicates PMT activation time: the darker blue color shows earlier first hit arrivals. The primary vertex is shown by the gray sphere}
	\label{fig:data_vis}
\end{figure*}

Figure~\ref{fig:data_vis} illustrates one example event for the positron with a \num{6.165}~MeV deposited energy, as seen by the 20-inch PMTs. The figure shows accumulated charges in PMT channels (left) and activation time, i.e. the first hit arrival times (right). 

In this study, event energy reconstruction is based only on the information collected by the larger 20-inch PMTs. We do not include the smaller ones (3-inch) so far because their contribution to the light collection is negligible.

To train models and to evaluate their performance, we prepared two datasets:
\begin{enumerate}
        
        \item \textbf{Training dataset} consisting of 5 million events uniformly spread in the scintillator volume with isotropic angular distribution and with a flat kinetic energy spectrum ranging from 0 to 10 MeV, i.e. $E_{\text{kin}} \in [0, 10]$ MeV.
        
        \item \textbf{Testing dataset} consisting of 14 subsets with discrete kinetic energies of 0~MeV, 0.1~MeV, 0.3~MeV, 0.6~MeV, 1~MeV, 2~MeV, ..., 10~MeV and with the same spatial and angular distribution as the training dataset. Each subset contains about 100 thousand events.
\end{enumerate}

\section{Machine Learning Approach}
\label{sec:ml}

\begin{figure}[H]
    \centering
    \includegraphics[width=1\columnwidth]{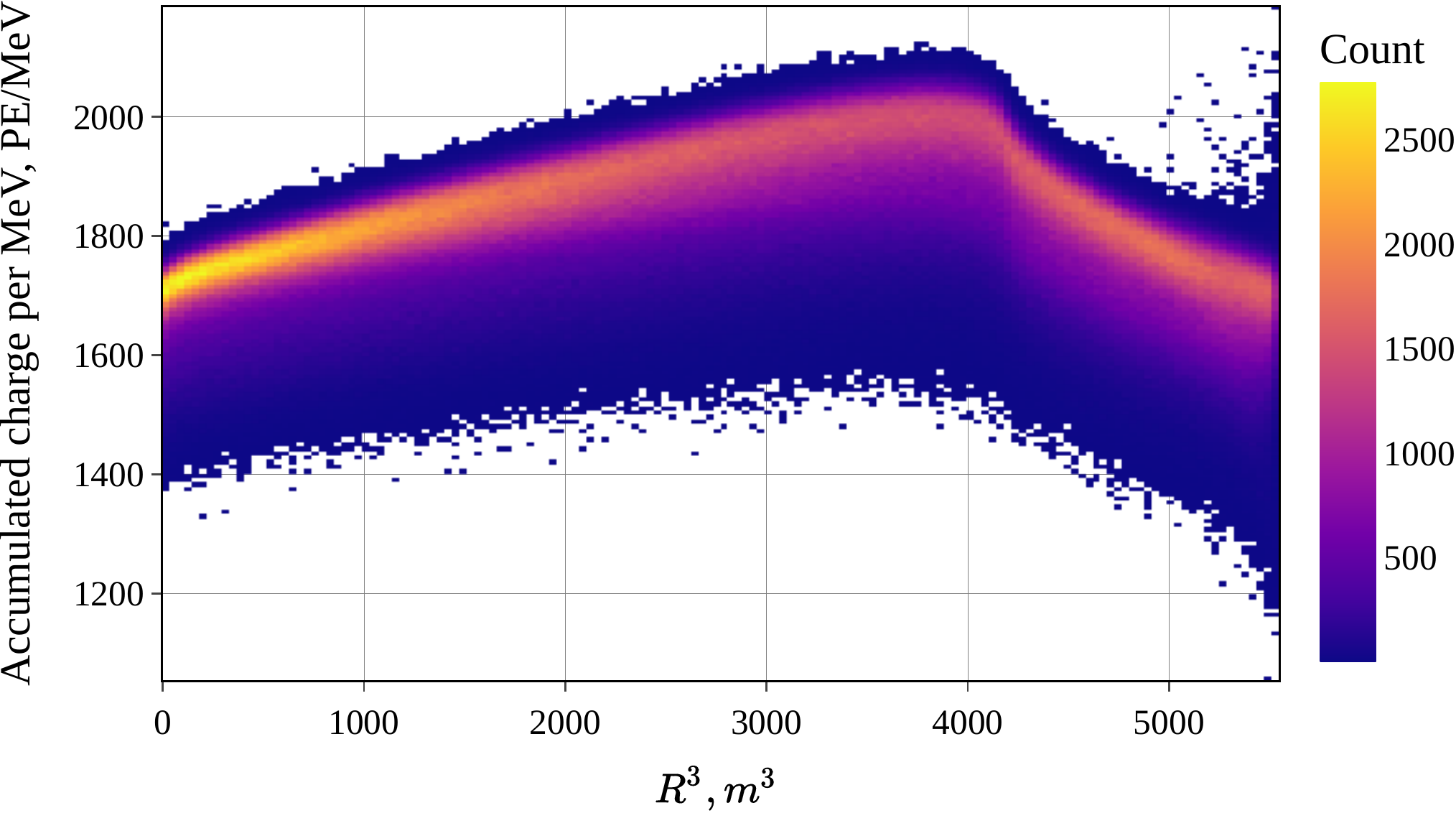}\\
    \caption{Accumulated charge on PMTs per 1 MeV of deposited energy as a function of the radius cubed $R^3$ for positrons}
    \label{fig:mean_charge_vs_R}
\end{figure}

The event energy defines the total charge collected by all PMTs because the number of emitted photons is roughly proportional to the deposited energy. However, the total collected charge also depends on the event position. This nonuniformity is illustrated in Figure~\ref{fig:mean_charge_vs_R} which shows the accumulated charge per 1~MeV of the deposited energy as a function of the radius cubed $R^3$. The sharp decrease of the accumulated charge in the region of $R$ $\gtrsim$ 16~m (or $R^3$ $\gtrsim$ 4000~m$^3$) is caused by the effect of the total internal reflection due to which the photons with a large incident angle never escape the target.
In the case of JUNO, spherical symmetry is not exactly held because the supporting structures at the bottom hemisphere hinder the installation of some PMTs, which leads to additional dependence on the $z$-coordinate. 

The pattern of signal distribution over PMTs carries the information on the event position and may be used for a precise reconstruction of energy taking into account nonuniformity.
Two numbers from each of the 17612 20-inch PMTs of the JUNO Central Detector can be extracted: the total charge in the event window and the first hit time (FHT), resulting in 35224 channels (some of which may be empty).

The values of FHT are the result of the waveform reconstruction and the trigger algorithms embedded in the JUNO simulation software. The procedure is the same as it will be used for the real data and does not rely on the true simulation information in any sense. 
After an event starts, the main part of the charge is collected by PMTs in about \num{200}~ns (depending on the distance from the center of the detector). In the remaining time of the signal window, there is a noticeable contribution from scattered and reflected photons as well as from dark noise.

In this study, we aggregate the information obtained channel-wise in a small set of features in order to reduce the number of channels. This set of features is used as an input to machine learning models to predict the deposited energy $E_{\text{dep}}$.

In this section, we describe how we construct features for machine learning algorithms and consider two models: Boosted Decision Trees and Fully Connected Deep Neural Network. The models are trained and tested on exactly the same datasets, which enables a fair comparison given in Section~\ref{sec:results}.

\subsection{Feature Engineering}
\label{sec:mlFeatEng}

        \begin{figure*}[!htb]
            \captionsetup[subfigure]{labelformat=empty}
            \centering
            \subfloat[]{\includegraphics[width=1\columnwidth]{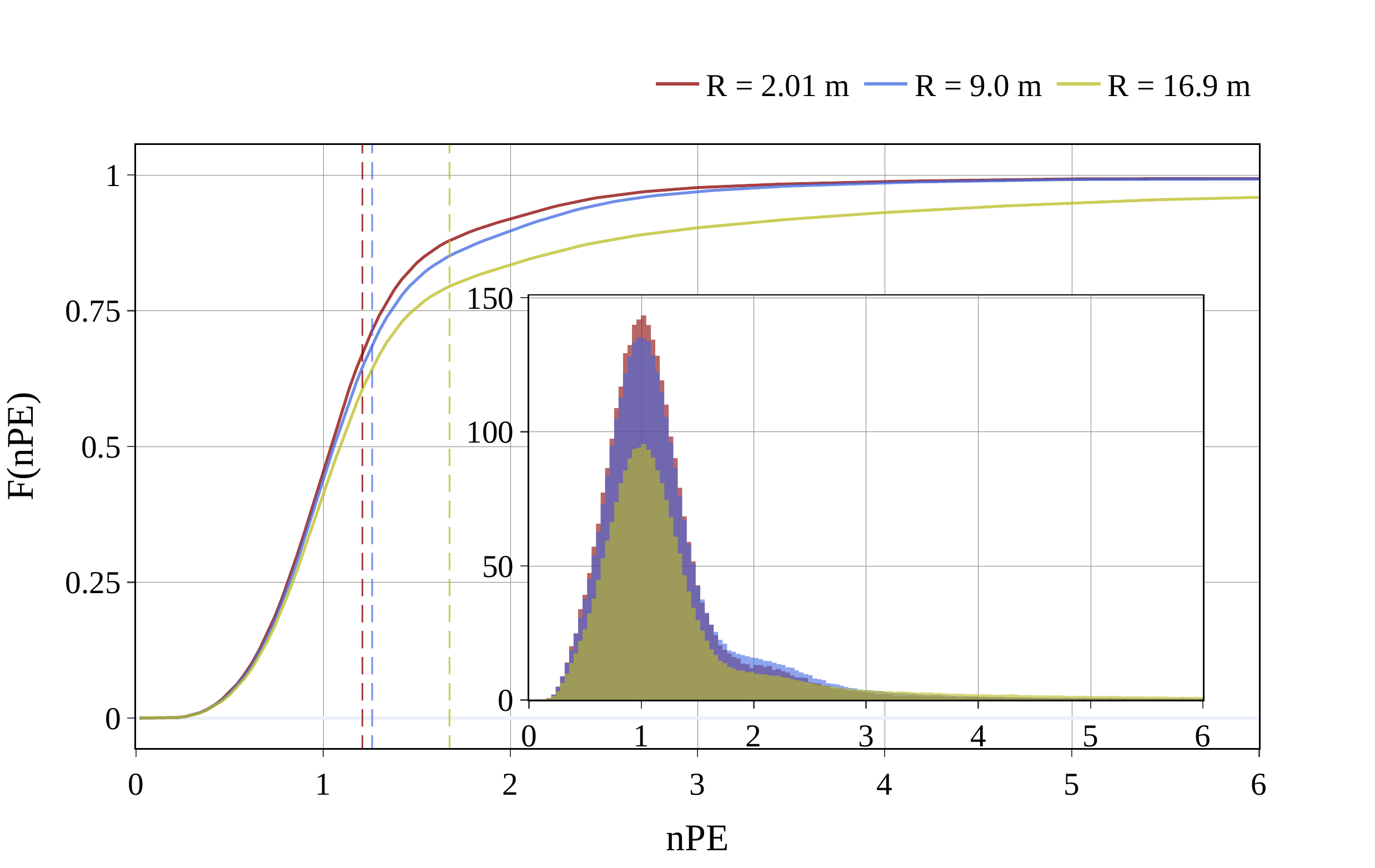}\label{fig:plot_cdf_pdf_nPE_R}}
            \subfloat[]{\includegraphics[width=1\columnwidth]{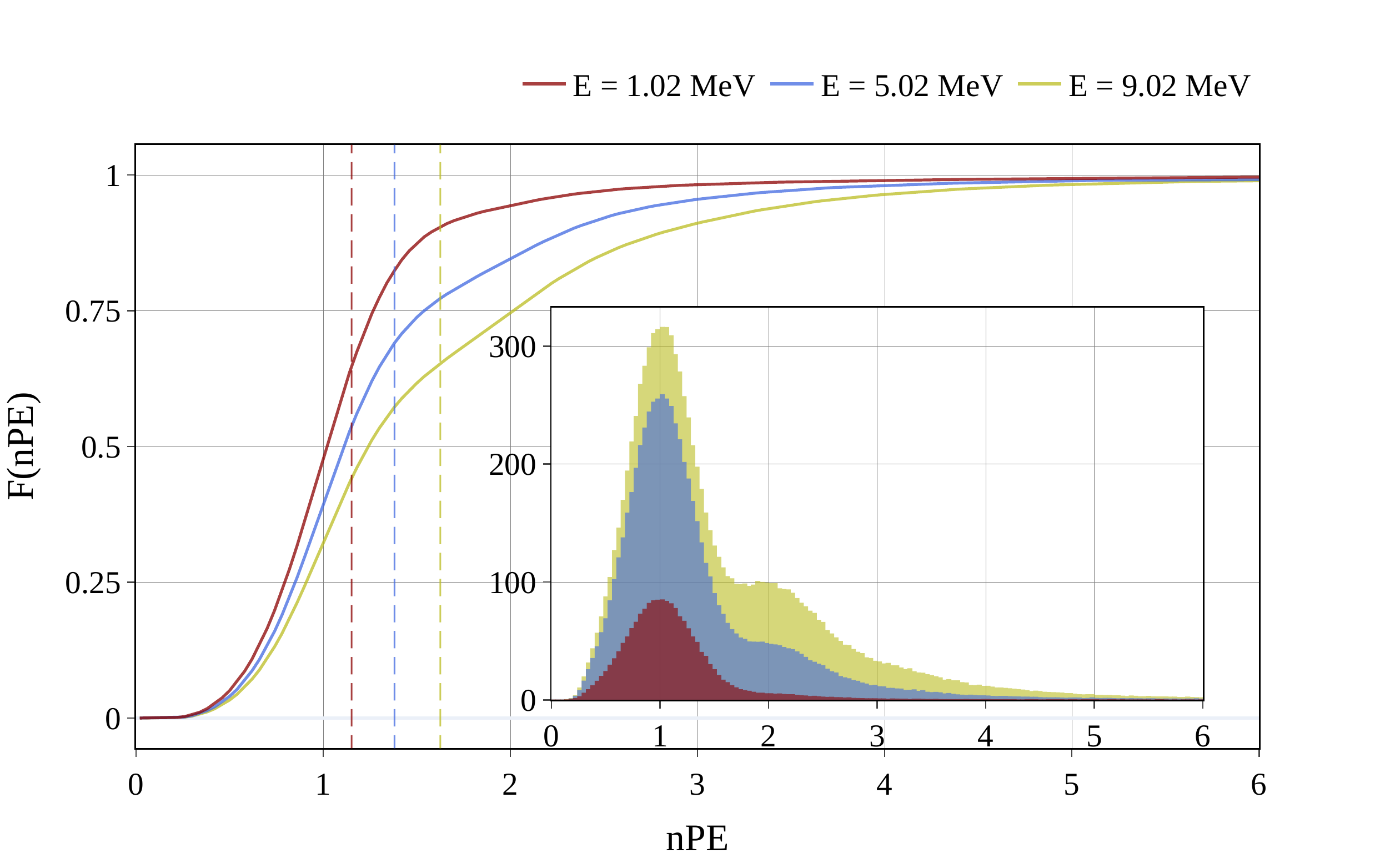}\label{fig:plot_cdf_pdf_nPE_E}}
            \caption{Examples of CDFs and PDFs for charge distributions for events with the kinetic energy of $E_{\rm kin} = 1$~MeV but different radius (left) and events at the center of the detector ($R=0$) but different energies (right). Dashed lines denote PDF mean values}
            \label{fig:npe_cdf_pdf}
        \end{figure*}

The following basic aggregated features are extracted for energy reconstruction:

\begin{enumerate}

    \item \texttt{AccumCharge} --- total charge accumulated on all fired PMTs. In the first order, it is proportional to $E_{\rm dep}$ and, therefore, is expected to be a powerful feature.
    
    \item \texttt{nPMTs} --- total number of fired PMTs. It is similar to \texttt{AccumCharge} but may bring some complementary information.

	\item Features aggregated from the charge and position of individual PMTs:
 
	\begin{enumerate}[label*=\arabic*.]
	    \item Coordinate components of the center of charge:
    	\begin{equation}
            \begin{split}
    	    (x_{\rm cc},\ y_{\rm cc},\ &z_{\rm cc}) = \vec{r}_{\rm cc} = \\ &=  \frac{\sum_{i=1}^{N_{\rm PMTs}} \vec{r}_{{\rm PMT}_i}\cdot n_{{\rm p.e.}, i}}{\sum_{i=1}^{N_{\rm PMTs}}n_{{\rm p.e.}, i}},
            \end{split}
    	\end{equation}
    	and its radial component: 
    	\begin{equation}
    	    R_{\rm cc} = |\vec{r}_{\rm cc}|
    	\end{equation}
     
        Coordinate components of the center of charge provide a rough approximation of the location of the energy deposition and help to correct the nonuniformity of the detector response. 
        It is useful, for some ML models, to engineer new synthesized features from the existing ones~\cite{Coates:2011, Heaton:2016}. It is sometimes hard, in particular for Boosted Decision Trees, to learn nonlinear dependencies, and so we construct the following extra features:
        \begin{align}
            \label{eq:synfs}
            \begin{split}
            \theta_{\rm cc} &= \arctan{\frac{\sqrt{x_{\rm cc}^2 + y_{\rm cc}^2}}{z_{\rm cc}}},\ 
            \phi_{\rm cc} = \arctan{\frac{y_{\rm cc}}{x_{\rm cc}}},\\
            J_{\rm cc} &= R_{\rm cc}^2 \cdot \sin{\theta_{\rm cc}},\  
            \rho_{\rm cc} = \sqrt{x_{\rm cc}^2 + y_{\rm cc}^2},\\
            \gamma_{z}^{\rm cc} &= \frac{z_{\rm cc}}{\sqrt{x_{\rm cc}^2 + y_{\rm cc}^2}},\
            \gamma_{y}^{\rm cc} = \frac{y_{\rm cc}}{\sqrt{x_{\rm cc}^2 + z_{\rm cc}^2}},\\
            \gamma_{x}^{\rm cc} &= \frac{x_{\rm cc}}{\sqrt{z_{\rm cc}^2 + y_{\rm cc}^2}}.
            \end{split}
        \end{align}    

        \item Shape of charge distribution over PMT channels. Figure~\ref{fig:npe_cdf_pdf} shows cumulative distribution functions (CDFs) and probability density functions (PDFs) for charge distribution on fired PMTs (nPE).
        The PDFs of nPE consist of two prominent peaks corresponding to one or two hits on a single PMT. The higher the energy, the more the second peak is populated. Moreover, for the events occurring near the edge there are PMTs accumulating a large amount of charge and forming a long tail of the distribution. Therefore, the shape of this distribution delivers information both on the intensity of light emission and its position. 
        To characterize the CDF for charge, we use the following sets of percentiles as features:
        $$
        \{\rm pe_{2\%}, \rm pe_{5\%}, \rm pe_{10\%}, \rm pe_{15\%}, ..., \rm pe_{90\%}, \rm pe_{95\%}\}
        $$
        as well as mean, standard deviation, skewness, and kurtosis:
        $$
            \{\rm pe_{mean}, \rm pe_{std}, \rm pe_{skew}, \rm pe_{kurtosis}\}
        $$
        \end{enumerate}

        \begin{figure*}[!htb]
            \captionsetup[subfigure]{labelformat=empty}
            \centering
            \subfloat[]{\includegraphics[width=1\columnwidth]{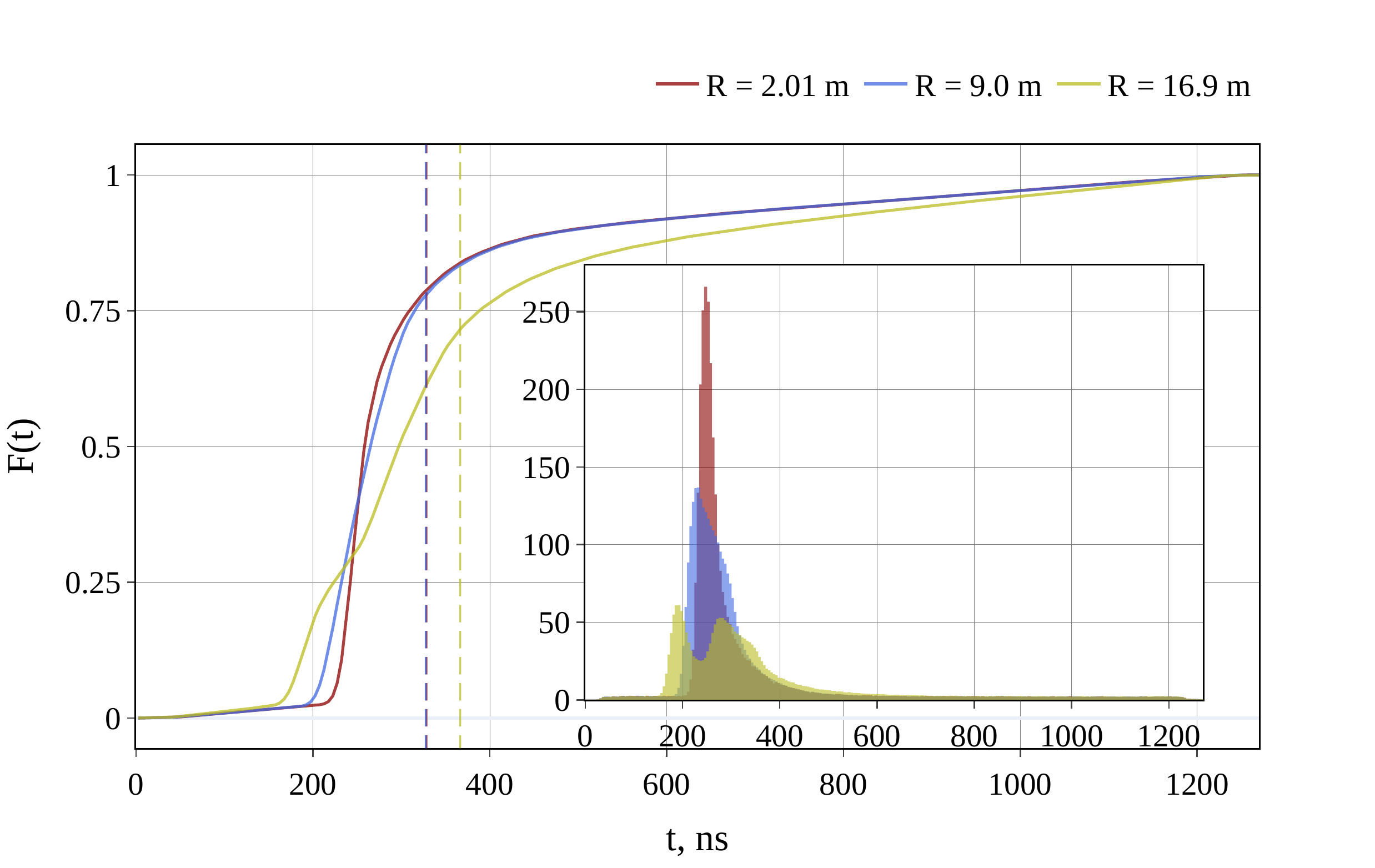}\label{fig:plot_cdf_pdf_FHT_R}}
            \subfloat[]{\includegraphics[width=1\columnwidth]{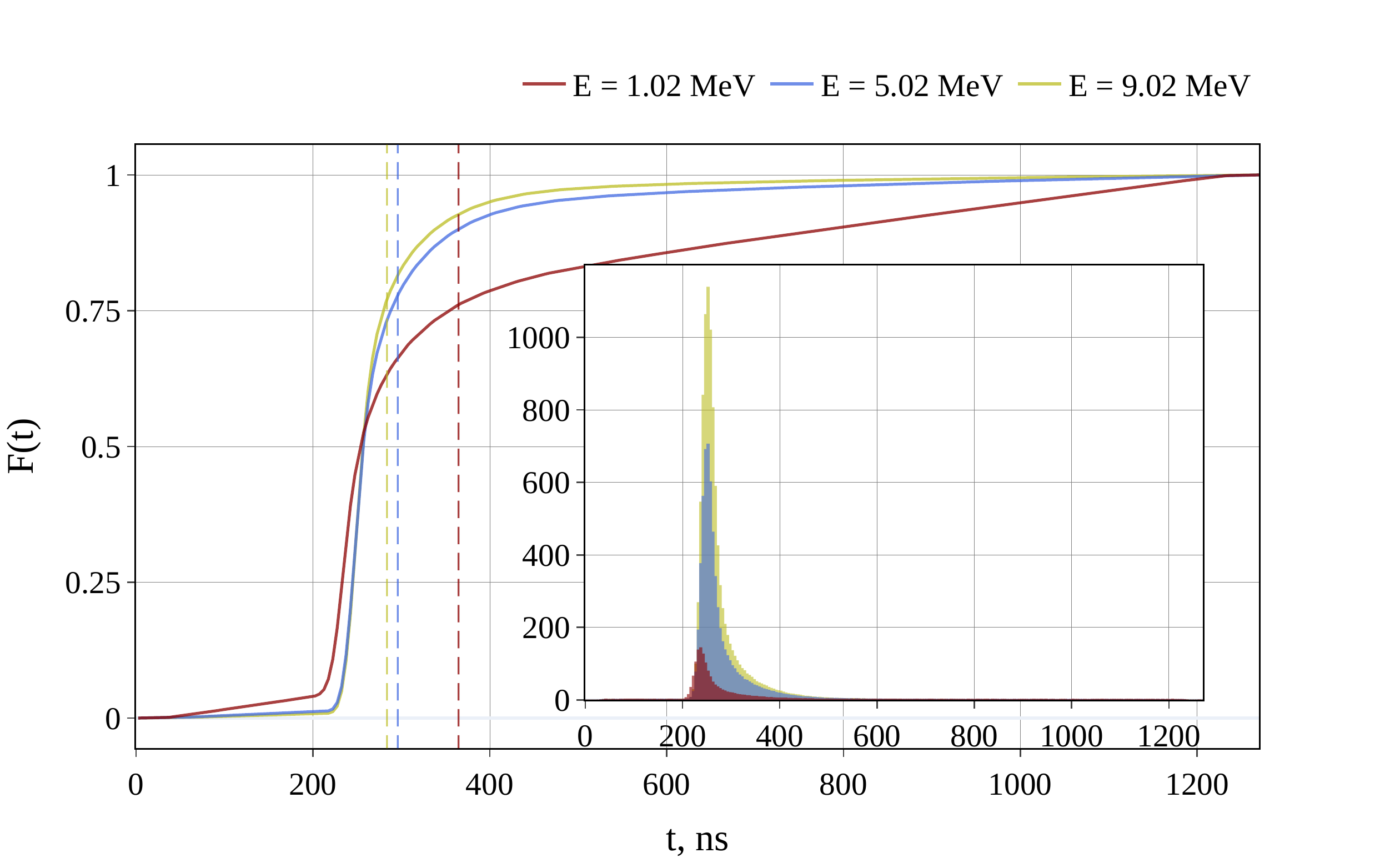}\label{fig:plot_cdf_pdf_FHT_E}}
            \caption{Examples of CDFs and PDFs for FHT distributions for events with the kinetic energy of $E_{\rm kin} = 1$~MeV but different radius (left) and events at the center of the detector ($R=0$) but different energies (right). Dashed lines denote PDF mean values}
            \label{fig:fht_cdf_pdf}
        \end{figure*}

	\item Features aggregated from first hit times and position of individual PMTs:
	\begin{enumerate}[label*=\arabic*.]
        \item Coordinate components of the FHT center:
    	\begin{equation}
            \begin{split}
    	    (x_{\rm cht},\ y_{\rm cht},\ &z_{\rm cht}) = \vec{r}_{\rm cht} = \\ &=\frac{1}{\sum_{i=1}^{N_{\rm PMTs}} \frac{1} {t_{{\rm ht},i} + c}} \sum_{i=1}^{N_{\rm PMTs}} \frac{\vec{r}_{{\rm PMT}_i}} {t_{{\rm ht},i} + c},
            \end{split}
            \end{equation}
    	and its radial component: 
    	\begin{equation}
    	    R_{\rm cht} = |\vec{r}_{\rm cht}|
    	\end{equation}
    	Here the constant $c$ is required to avoid division by zero. The value of 50 ns was selected to make the center of FHT closer to the energy deposition vertex. These features give extra information on the location of the energy deposition.
        Likewise the center of charge, the center of FHT provides a rough estimation of the energy deposition center but with the use of time information. Figure~\ref{fig:R_cht_vs_R_cc} shows that $R_{\rm cht}$ and $R_{\rm cc}$ bring complementary information, especially at larger radii.
        
        \begin{figure}[!htb]
            \centering
            \includegraphics[width=0.75\columnwidth]{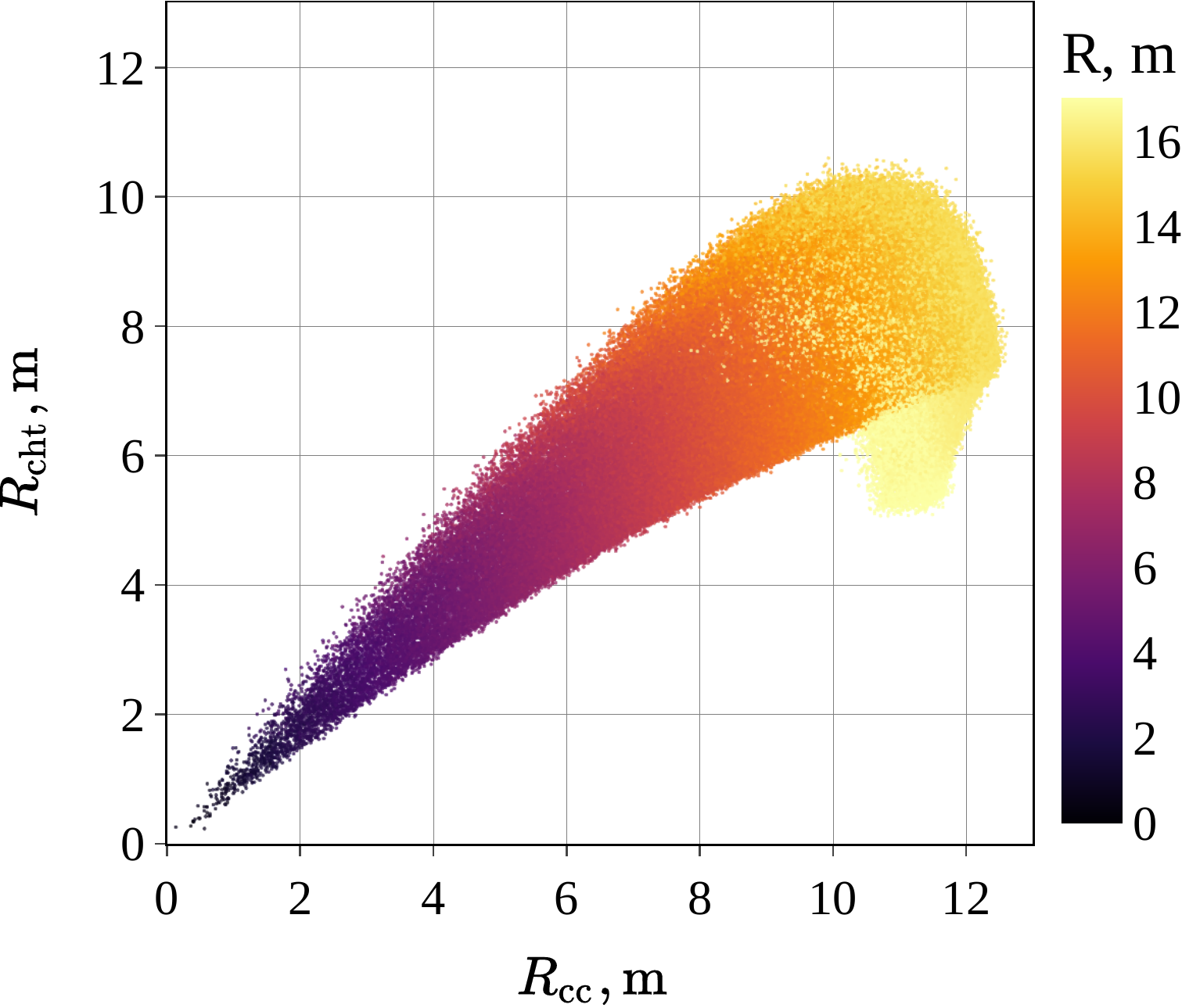}\\
            \caption{Correlation between $R_{\rm cc}$ and $R_{\rm cht}$ and the corresponding values of the radial component of energy deposition $R$ represented in color}
            \label{fig:R_cht_vs_R_cc}
        \end{figure}
        
        We also synthesize new features from the components of the center of FHT as it was done for the center of charge \eqref{eq:synfs}:
        \begin{align}
            \begin{split}
            \theta_{\rm cht} &= \arctan{\frac{\sqrt{x_{\rm cht}^2 + y_{\rm cht}^2}}{z_{\rm cht}}},\ 
            \phi_{\rm cht} = \arctan{\frac{y_{\rm cht}}{x_{\rm cht}}},\\
            J_{\rm cht} &= R_{\rm cht}^2 \cdot \sin{\theta_{\rm cht}},\  
            \rho_{\rm cht} = \sqrt{x_{\rm cht}^2 + y_{\rm cht}^2},\\
            \gamma_{z}^{\rm cht} &= \frac{z_{\rm cht}}{\sqrt{x_{\rm cht}^2 + y_{\rm cht}^2}},\
            \gamma_{y}^{\rm cht} = \frac{y_{\rm cht}}{\sqrt{x_{\rm cht}^2 + z_{\rm cht}^2}},\\
            \gamma_{x}^{\rm cht} &= \frac{x_{\rm cht}}{\sqrt{z_{\rm cht}^2 + y_{\rm cht}^2}}.
            \end{split}
        \end{align}

        \item        
        Shape of FHT distribution over PMT channels. The corresponding PDFs and CDFs are shown in Figure~\ref{fig:fht_cdf_pdf}.
        There is a clear dependence of the shape on both the energy and radial position. The depression of FHT PDFs at 16.9~m (yellow in the left panel of Figure~\ref{fig:fht_cdf_pdf}) is due to the total internal reflection of photons that have large incidence angles at the spherical surface of the detector. Near the edge, the events appear earlier in the readout window because of the lower time-of-flight correction. The FHT PDFs at the detector center (right panel of Figure~\ref{fig:fht_cdf_pdf}) have similar shapes only differing by the relative contribution from dark noise hits and from annihilation gammas. This contribution decreases with increasing energy.
        To characterize the CDF for FHT, we add the following percentile features as we did for charge:
        $$
            \{\rm ht_{2\%}, \rm ht_{5\%}, \rm ht_{10\%}, \rm ht_{15\%}, ..., \rm ht_{90\%}, \rm ht_{95\%}\}
        $$
        as well as mean, standard deviation, skewness and kurtosis:
        $$
            \{ \rm ht_{mean}, \rm ht_{std}, \rm ht_{skew}, \rm ht_{kurtosis}\}
        $$
        Figure~\ref{fig:plot_avg_signal_FHT} shows the average PDF for FHT distribution for 161 events with $E_{\rm dep} = 2.022$~MeV and $R = 16$~m. There is a noticeable variance in the first bins. Taking this into account, we also use the following differences between percentiles, since they are more robust to this variance: 
            $$\{\rm ht_{5\% - 2\%}, \rm ht_{10\% - 5\%}, ..., \rm ht_{95\% - 90\%}\}$$
        
        \begin{figure}[!htb]
        	\centering
        	\includegraphics[width=1\columnwidth]{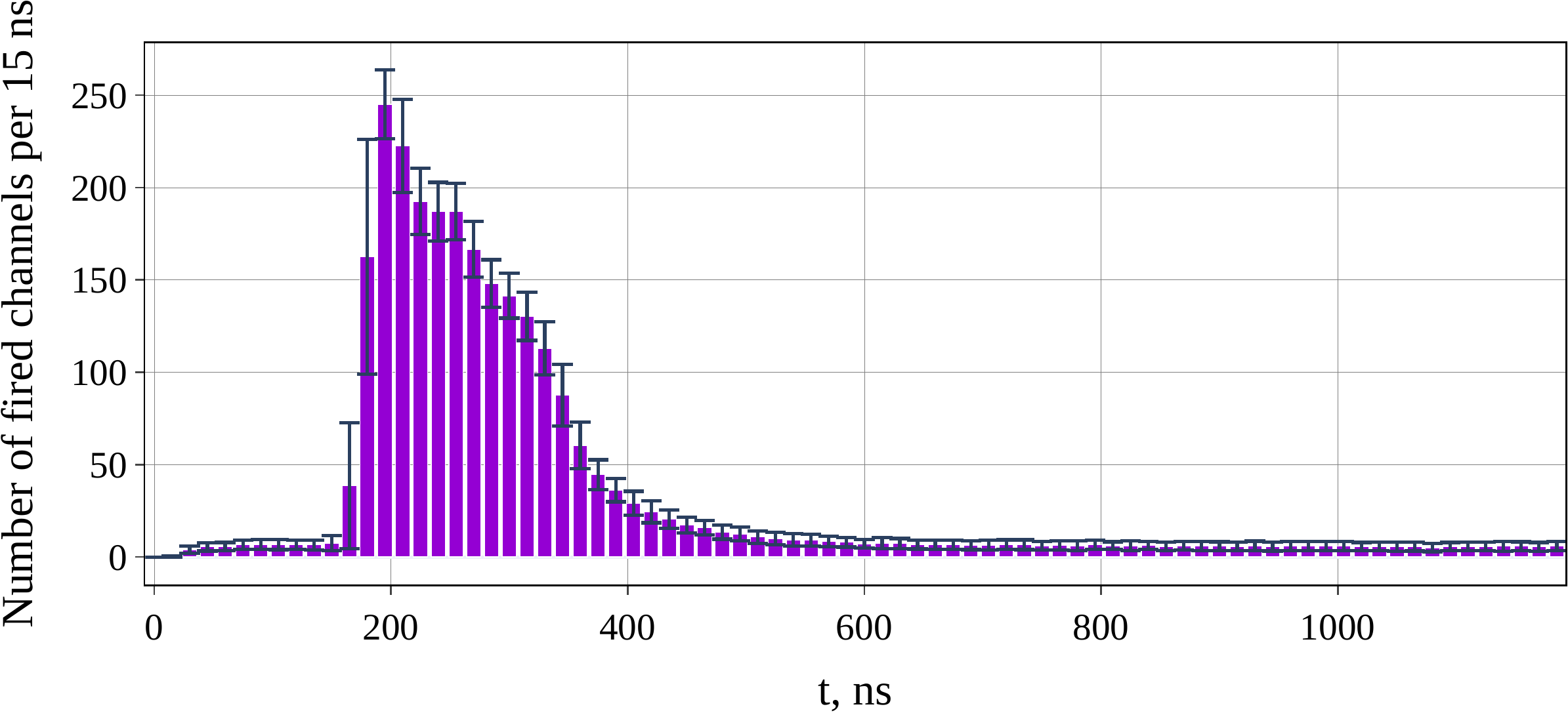}\\
        	\caption{Average PDF for FHT distribution with deposited energy $E_{\rm dep} = 2.022$~MeV and $R = 16$~m. Error bars represent the standard deviation in each corresponding bin}
        	\label{fig:plot_avg_signal_FHT}
        \end{figure}
    \end{enumerate}
\end{enumerate}    

\subsubsection{Feature Summary}

In total, we engineered 91 features. The full set of features is listed in Table~\ref{tab:features_list}.

\begin{table*}[!htb]
    \centering
    \begin{tabular}{|l|l|}
        \hline
        \it{Notation} & \it{Description} \\
        \hline
        \texttt{AccumCharge} & Total accumulated charge \\
        \hline
        \texttt{nPMTs}       & Number of fired PMTs \\
        \hline
        $x_{\rm cc}$, $y_{\rm cc}$, $z_{\rm cc}$, $R_{\rm cc}$, $\theta_{\rm cc}$, $\phi_{\rm cc}$, $J_{\rm cc}$, $\rho_{\rm cc}$, $\gamma_{z}^{\rm cc}$, $\gamma_{y}^{\rm cc}$, $\gamma_{x}^{\rm cc}$ 
                             & Center of charge \\
        \hline
        $\rm pe_{n\%}$, $\rm pe_{mean}$, $\rm pe_{std}$, $\rm pe_{skew}$, $\rm pe_{kurtosis}$
                             & Charge distribuion \\
        \hline
        $x_{\rm cht}$, $y_{\rm cht}$, $z_{\rm cht}$, $R_{\rm cht}$, $\theta_{\rm cht}$, $\phi_{\rm cht}$, $J_{\rm cht}$, $\rho_{\rm cht}$, $\gamma_{z}^{\rm cht}$, $\gamma_{y}^{\rm cht}$, $\gamma_{x}^{\rm cht}$ 
                             & Center of FHT \\
        \hline
        $\rm ht_{n\%}$, $\rm ht_{n_{i+1}\% - n_i\%}$, $\rm ht_{mean}$, $\rm ht_{std}$, $\rm ht_{skew}$, $\rm ht_{kurtosis}$
                             & FHT distribution \\
        \hline
    \end{tabular}
    \caption{List of all feature notations with brief descriptions. Here, n $\in$ \{2, 5, 10, 15, ..., 90, 95\}}
    \label{tab:features_list}
\end{table*}

\subsection{Boosted Decision Trees}
\label{sec:mlBDT}

In terms of machine learning, the problem of energy reconstruction is a supervised regression problem. In general, algorithms for supervised problems learn the mapping of input features to a target output using a data sample with input-output pairs. Boosted Decision Trees (BDT) is one of such algorithms~\cite{Friedman:2002, Friedman:2001}.

BDT is a gradient boosting-based algorithm with Decision Tree (DT)~\cite{Quinlan:1987} as a base model.
DT is a simple, fast, and interpretable model.
DT consists of a binary set of splitting rules based on values of different features of the object.
A single DT is not a powerful algorithm. Therefore, an ensemble of DTs is commonly used in which DTs are trained sequentially: each subsequent DT is trained to correct errors of previous DTs in the ensemble.

In general, boosting is an algorithm of sequential combining of some basic models to create a stronger one. There are many different realisations of boosting algorithms. For example, AdaBoost~\cite{AdaBoost}, widely used in the past, is based on the weighting procedure: a weight is assigned to each event at each iteration. Events with the worse prediction get a larger weight, and the ones with the better prediction get a smaller weight. In contrast, gradient boosting defines a differentiable loss function and calculates the gradient between the true targets ($E_{\rm dep}$) and the ones predicted by the previous ensemble. Then, at each iteration, the parameters of the next basic model are updated to fit the residual gradient from the previous ensemble.

Tree-based models, including BDT, are an efficient way to work with tabular data~\cite{Borisov:2021}. 
XGBRegressor from the XGBoost library~\cite{Chen:2016} is adopted in this paper. XGBoost is a robust and widely-accepted framework for training and deployment of BDT.

\subsubsection{Feature Selection}
\label{sec:mlBDTFeatSel}

\begin{figure}[!htb]
	\centering
 	\includegraphics[width=1\columnwidth]{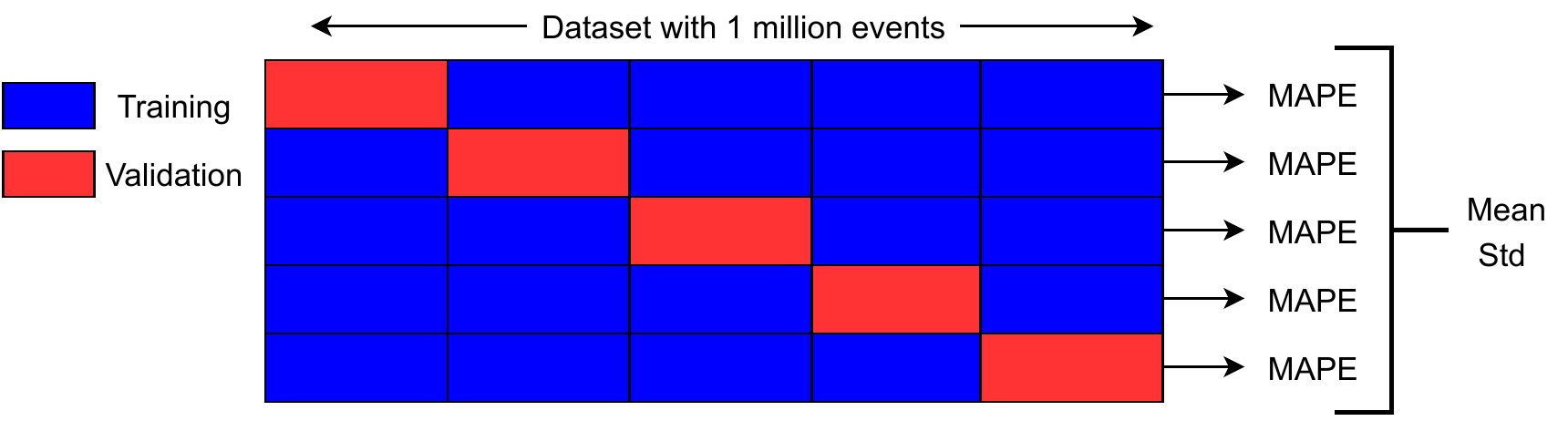}
 	\caption{5-fold cross-validation procedure}
 	\label{fig:CV}
 \end{figure}

Many of the 91 features described in Section~\ref{sec:mlFeatEng} are highly correlated, and we want to keep only a subset of features which provide the same performance of the model as the full set. 
The feature selection procedure for the BDT model is described as follows:
\begin{enumerate}
    \item We train the BDT model on a dataset with 1 million events using a 5-fold cross-validation from the Scikit-learn library~\cite{Scikit-learn:2011}.
    Figure~\ref{fig:CV} illustrates the cross-validation with the early stopping condition on an additional hold-out dataset.
    We evaluate the model performance with a mean absolute percentage error (MAPE), which is better suited for reconstruction at low energies. MAPE is a generalized metric that is calculated for true deposited energies and reconstructed deposited energies for, in case of training, validation dataset: 
    \begin{equation}
        \rm{MAPE} = \frac{100\%}{N} \sum_{i=1}^{N} \left| \frac{y_i - \hat{y}_i}{y_i} \right|,
    \end{equation}
    where $N$ --- number of events, $y_i$ --- true deposited energy, $\hat{y}_i$ --- reconstructed deposited energy.
    As a result of the cross-validation procedure, we obtain the mean MAPE on the validation datasets and its standard deviation $\varepsilon$. 
    
    \item Then, we initialize an empty list and start filling it with features. At each step, we pick a feature from the full set and train the BDT model using the 5-fold cross-validation procedure with an early stopping.
    Then, we calculate the mean MAPE for each step and add to the list the feature that provides the stepwise best performance.
    We stop when the MAPE score differs from the mean MAPE score for the model trained on all features by less than the standard deviation $\varepsilon$. 

    \begin{figure}[!htb]
     	\centering
    	\includegraphics[width=1\columnwidth]{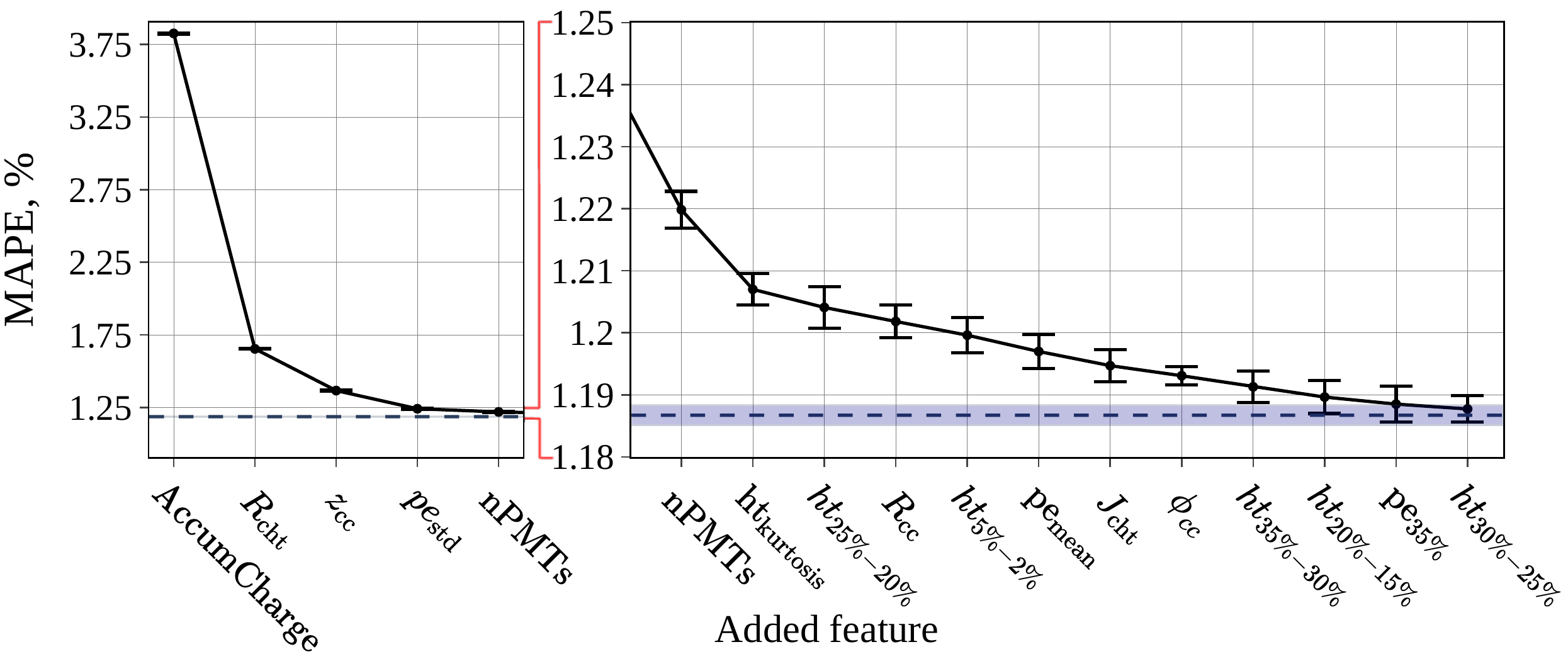}
    	\caption{Results of the feature selection procedure. Dashed line --- average MAPE for BDT trained on all features after CV with its standard deviation}
    	\label{fig:FeatureSelection}
    \end{figure}

\end{enumerate} 

Figure~\ref{fig:FeatureSelection} shows the results of the above-described feature selection procedure.
The procedure results in the following set of features:
\begin{multicols}{2}
\begin{enumerate}[label=\textbf{\arabic*})]
    \item \texttt{AccumCharge}
    \item $R_{\rm cht}$
    \item $z_{\rm cc}$
    \item $\rm pe_{std}$
    \item \texttt{nPMTs}
    \item $\rm ht_{kurtosis}$ 
    \item $\rm ht_{25\%-20\%}$  
    \item $R_{\rm cc}$
    \item $\rm ht_{5\%-2\%}$
    \item $\rm pe_{mean}$
    \item $J_{\rm cht}$
    \item $\phi_{\rm cc}$
    \item $\rm ht_{35\%-30\%}$
    \item $\rm ht_{20\%-15\%}$
    \item $\rm pe_{35\%}$
    \item $\rm ht_{30\%-25\%}$
\end{enumerate}
\end{multicols}

As expected, the most powerful feature is the total accumulated charge because the amount of emitted light is, to the first order, proportional to the deposited energy. However, the amount of detected photons is position-dependent, which can be corrected using the information provided by the coordinates of the center of charge and the center of FHT. Due to the axial symmetry of the detector, $x$ and $y$ components provide useful information only as a combination $x^2+y^2$, so they do not appear in the list at all. Extra information is extracted from the width of the p.e. number distribution which is peaked around 1 for central events and is populated with large values for peripheral events. The shape of the time profile also helps, but its contribution is small because of a large event-to-event variation.

\begin{figure}[!htb]
 	\centering
	\includegraphics[width=1\columnwidth]{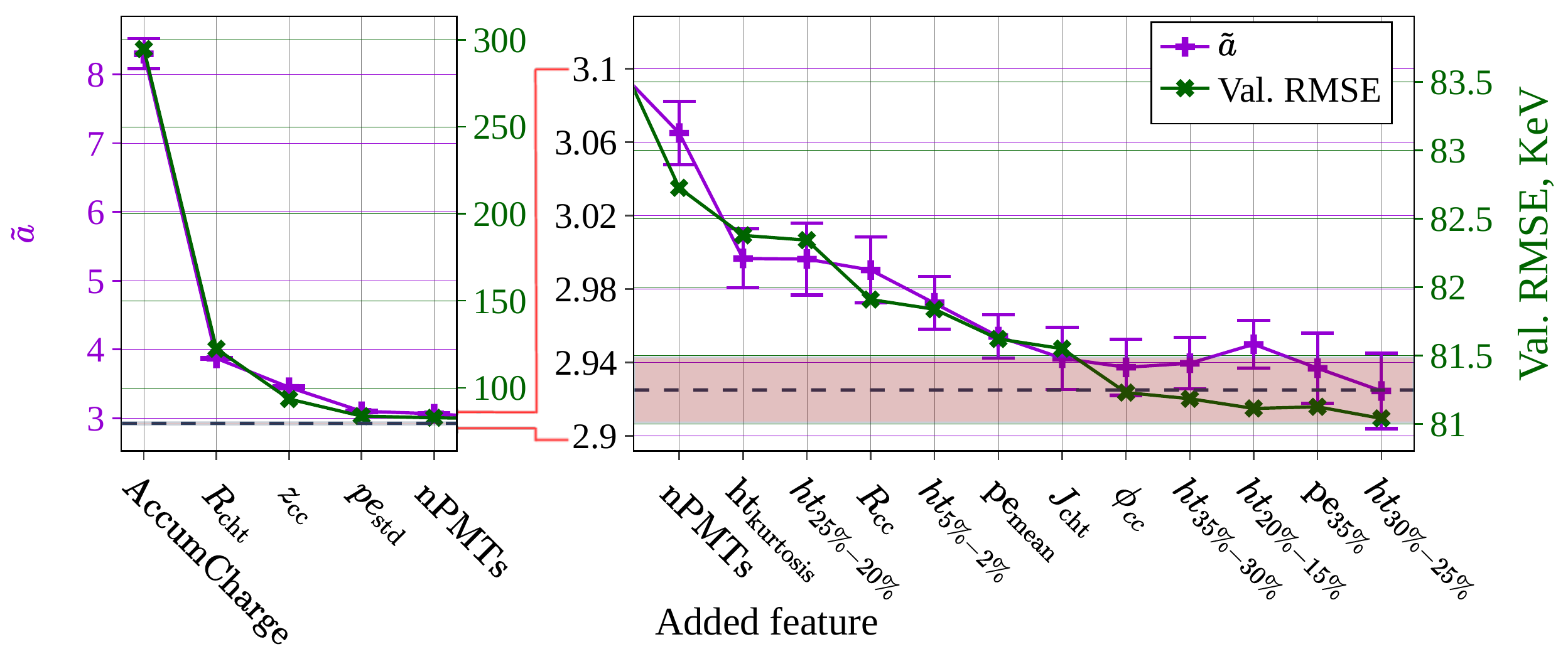}
	\caption{Dependence of the effective resolution $\tilde{a}$ and RMSE on the validation dataset for BDT on sequentially selected features of the model}
	\label{fig:a_tilde_ValRMSE}
\end{figure}

Figure~\ref{fig:a_tilde_ValRMSE} shows the dependence of effective resolution $\tilde{a}$ and standard deviation (RMSE)
obtained on the validation dataset for reconstructed energy and true energy. Both quantities are obtained with sequentially selected features as an input of the BDT model. RMSE is defined as follows:
\begin{equation}
\label{eq:rmse}
\rm{RMSE} = \sqrt{\frac{1}{N} \sum_{i=1}^{N} {(y_i - \hat{y}_i)^2}}.
\end{equation}
The dashed line illustrates $\tilde{a}$ for BDT trained on the full set of features, and the dark red area is its error.
The decrease of RMSE is consistent and converged with the improvement of the effective energy resolution $\tilde{a}$.
All models were trained with a maximal tree depth equal to 9 and a learning rate of 0.08, and with an early stopping condition with a patience of 5. 

\subsubsection{Hyperparameter Optimization}
\label{sec:mlBDTHypOpt}

We use the grid search approach for hyperparameter optimization for the BDT model.
The space of hyperparameters is determined, and the metric is calculated for each possible combination of hyperparameters from this space.
The node with the best metric is selected as the optimum. 

For the grid search, we use the BDT model with a learning rate equal to 0.08 trained with a 5-fold cross-validation on the dataset with 1 million events. We optimize the maximal tree depth in the ensemble. The training of the model stops if the MAPE score on the hold-out validation dataset has not decreased during 5 successive iterations.

\begin{figure}[!htb]
	\centering
 	\includegraphics[width=1\columnwidth]{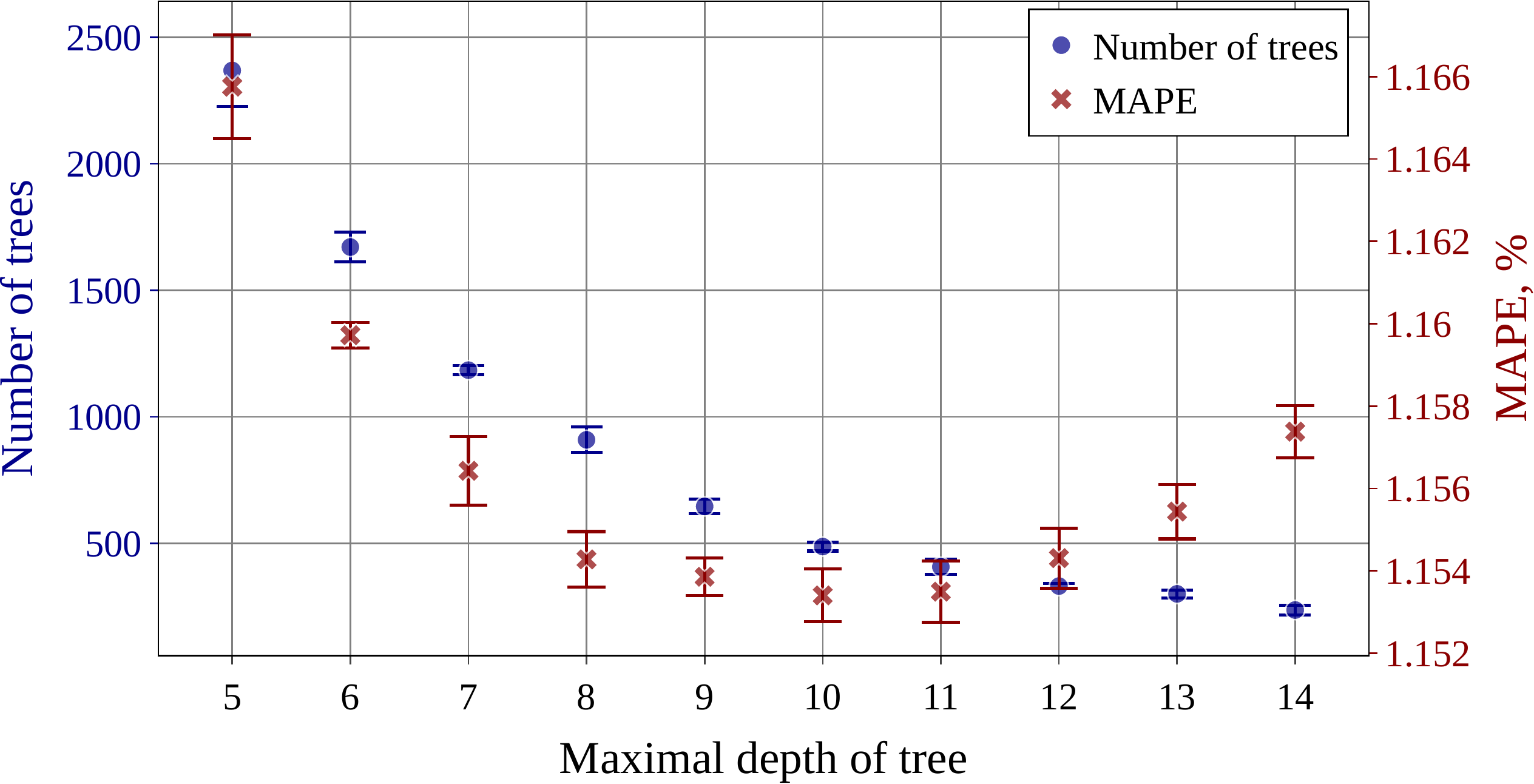}
 	\caption{Dependence of the minimal optimum number of trees and the corresponding MAPE score for the validation dataset on the maximum depth of a tree}
 	\label{fig:BDT_opt}
 \end{figure}

Figure~\ref{fig:BDT_opt} shows the results of hyperparameter optimization.
The best maximum depth of the tree was found to be 10, and the corresponding number of trees is 583. 
 
\subsection{Fully Connected Deep Neural Network}
\label{sec:mlFCDNN}

Training a neural network means fitting, based on data, the best parameters for mapping input features to output value(s). A fully connected neural network consists of layers with sets of units called neurons. Each neuron in a layer is connected with each neuron in the next layer. If it has many layers, it is called a Fully Connected Deep Neural Network (FCDNN).

Each neuron computes a linear combination of its inputs with its own learnable weights. To enable a network to reproduce nonlinearities, it is required to pass a neuron output through some nonlinear function, the so-called activation function. There are several popular choices: sigmoid, hyperbolic tangent, Rectified Linear Unit (ReLU) and its modifications (ELU, Leaky ReLU, SELU, etc.), and others~\cite{ActFuncs1, ActFuncs2}.

\begin{table*}[!htb]
	\centering
    \begin{tabular}{|l@{\hskip 60pt}|l|l@{\hskip 40pt}|}
    \hline
        \it{Hyperparameter} & \it{Range} & \it{Selected} \\
    \hline
	Units in input layer & [1, 512] & \textbf{256} \\
    \hline
	Units in hidden layers & [1, 512] & \textbf{256} \\    \hline
	Number of hidden layers & [1, 32] & \textbf{16} \\
    \hline
	Activation~\cite{ReLU, ELU, SELU} & ReLU, ELU, SELU & \textbf{ReLU} \\
    \hline
	Optimizer~\cite{Adam, SGD_RMSprop} & Adam, SGD, RMSprop & \textbf{Adam} \\
    \hline
	Learning rate & [0.0001, 0.01] & \textbf{0.0016} \\
    \hline
        Scheduler type~\cite{Exponential} & Exponential, None & \textbf{Exponential} \\
    \hline
	\vspace{-5pt} Input layer weights initialization  & & \\
	\vspace{-5pt} & normal, lecun-normal, uniform & \textbf{normal} \\
	Hidden layers weights initialization  & &  \\
    \hline
	\end{tabular}
	\caption{Hyperparameter search space for FCDNN. Selected hyperparameters are highlighted in bold}
	\label{tab:FCDNN_hp}
\end{table*}

Optimization of hyperparameters for FCDNN is performed using the BayesianOptimization tuner from the KerasTuner library for Python~\cite{kerastuner}.
To train the model, we use TensorFlow~\cite{tensorflow}. The
MAPE loss for reconstructed energy and true energy is used as a loss function. All input features were normalized with a standard score normalization.
The training process is performed with an early stopping condition on the validation dataset with a patience of 25 and with the batch size 1024. Table~\ref{tab:FCDNN_hp} shows the search space and the selected hyperparameters.

We first select several subsets of features that are reasonable from our previous experience. 
Then, we pick the following subset of features that provides virtually the same quality of the FCDNN model as the full set of features:
\begin{multicols}{2}
\begin{enumerate}[label=\textbf{\arabic*})]
    \item \texttt{AccumCharge}
    \item \texttt{nPMTs}
    \item $R_{\rm cc}$
    \item $R_{\rm cht}$
    \item $\rho_{\rm cc}$
    \item $\rho_{\rm cht}$
    \item $\rm pe_{mean}$
    \item $\rm pe_{std}$  
    \item $\rm pe_{skew}$
    \item $\rm pe_{kurtosis}$
\end{enumerate}
\end{multicols}
\begin{enumerate}[label=\textbf{\arabic*}), parsep=-10pt]
    \setcounter{enumi}{10}
    \item FHT distribution percetiles: \{$\rm ht_{2\%}$, $\rm ht_{5\%}$, $\rm ht_{10\%}$, $\rm ht_{15\%}$, ..., $\rm ht_{90\%}$, $\rm ht_{95\%}$\}
\end{enumerate}

Taking into account the axial symmetry, it is enough to have a pair of coordinates as features. We use $\rho = \sqrt{x^2 + y^2}$ and $R = \sqrt{x^2+y^2+z^2}$, however, the neural network can extract necessary information also from other combinations, e.g. $\rho$ and $z$.

It is interesting to note that with those features selected for the BDT approach, the prediction quality of FCDNN  is within $2\sigma$ from the optimal quality corresponding to the feature list above. 

\begin{figure}[!htb]
	\centering
	\includegraphics[width=1\columnwidth]{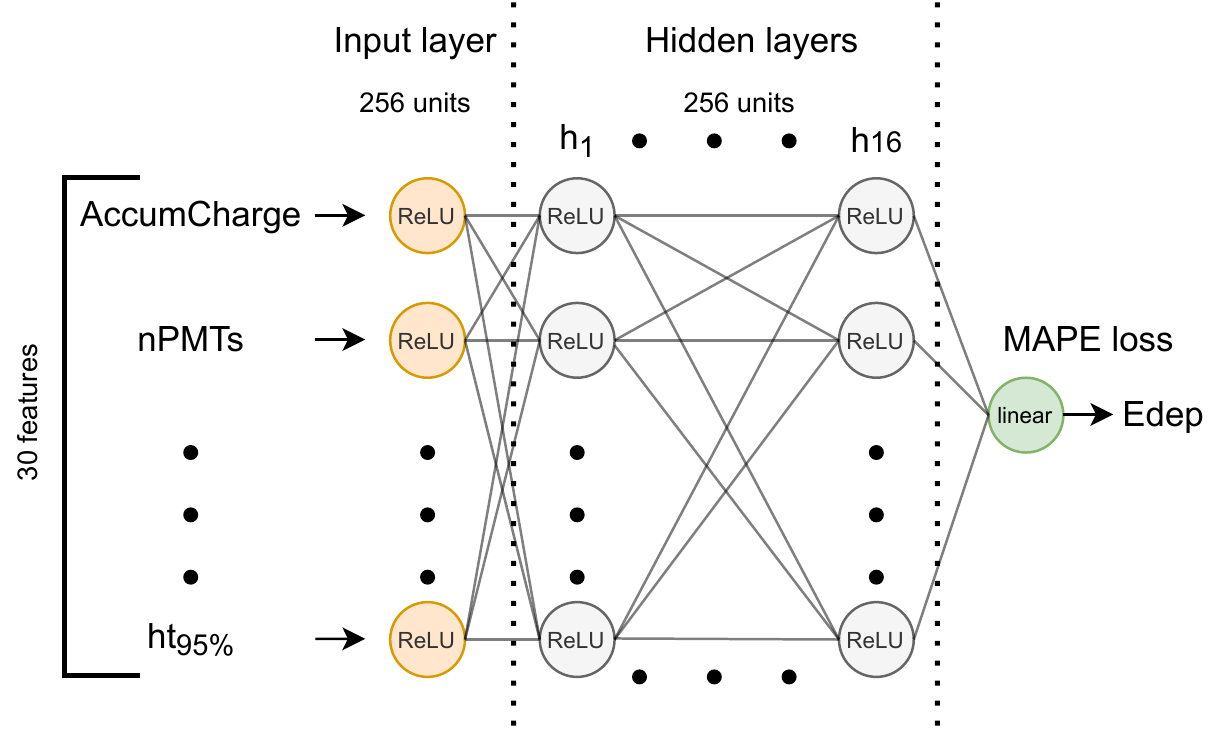}
	\caption{Selected FCDNN architecture and its main selected hyperparameters}
	\label{fig:FCDNN}
\end{figure}

Figure~\ref{fig:FCDNN} describes the selected architecture of FCDNN and its main selected hyperparameters.

\section{Results}
\label{sec:results}

\begin{figure*}[!htb]
	\centering
	\subfloat[]{\includegraphics[width=1\columnwidth]{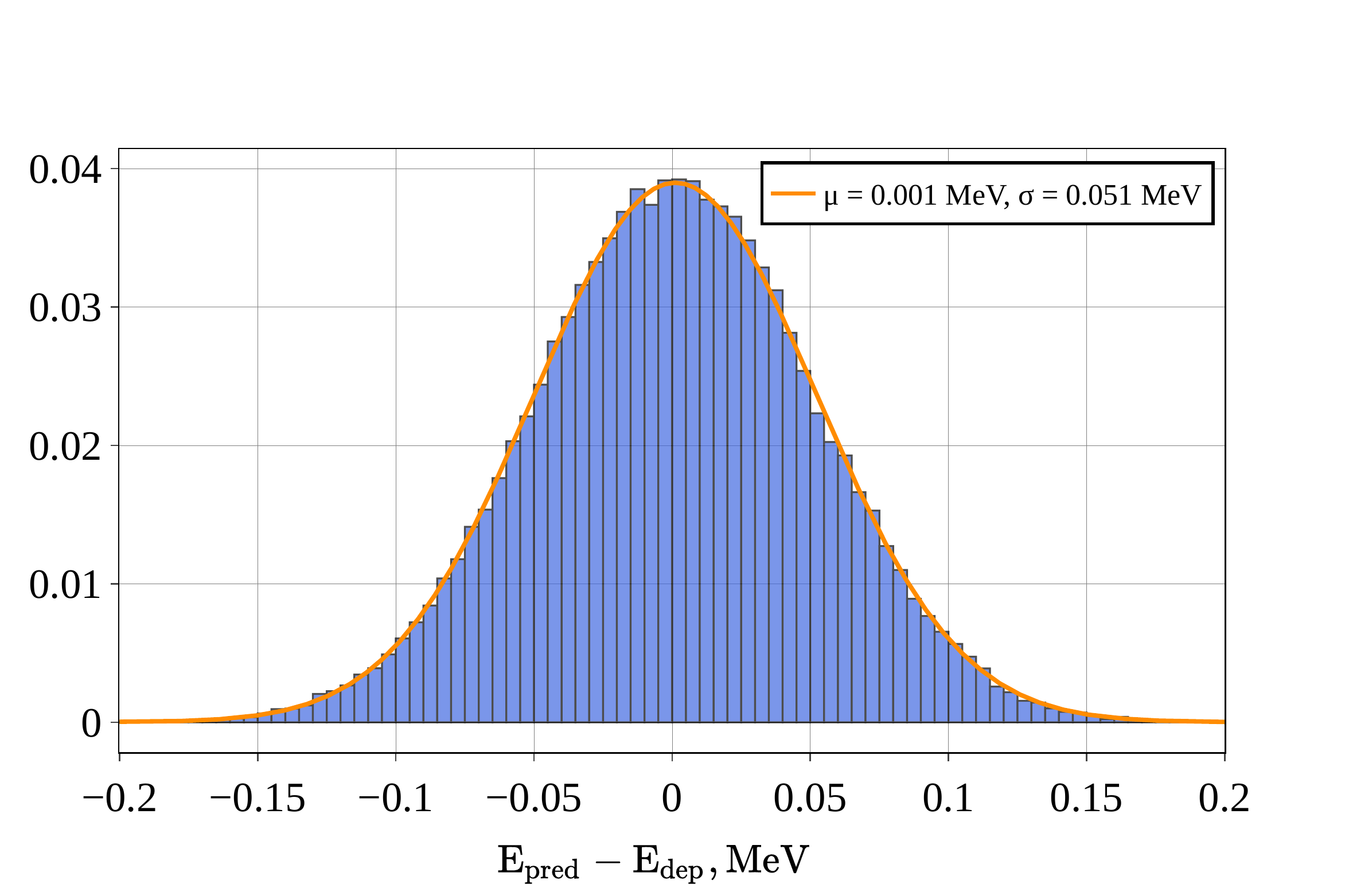}\label{fig:preds_distribution_BDT_3_022_MeV}}
	\subfloat[]{\includegraphics[width=1\columnwidth]{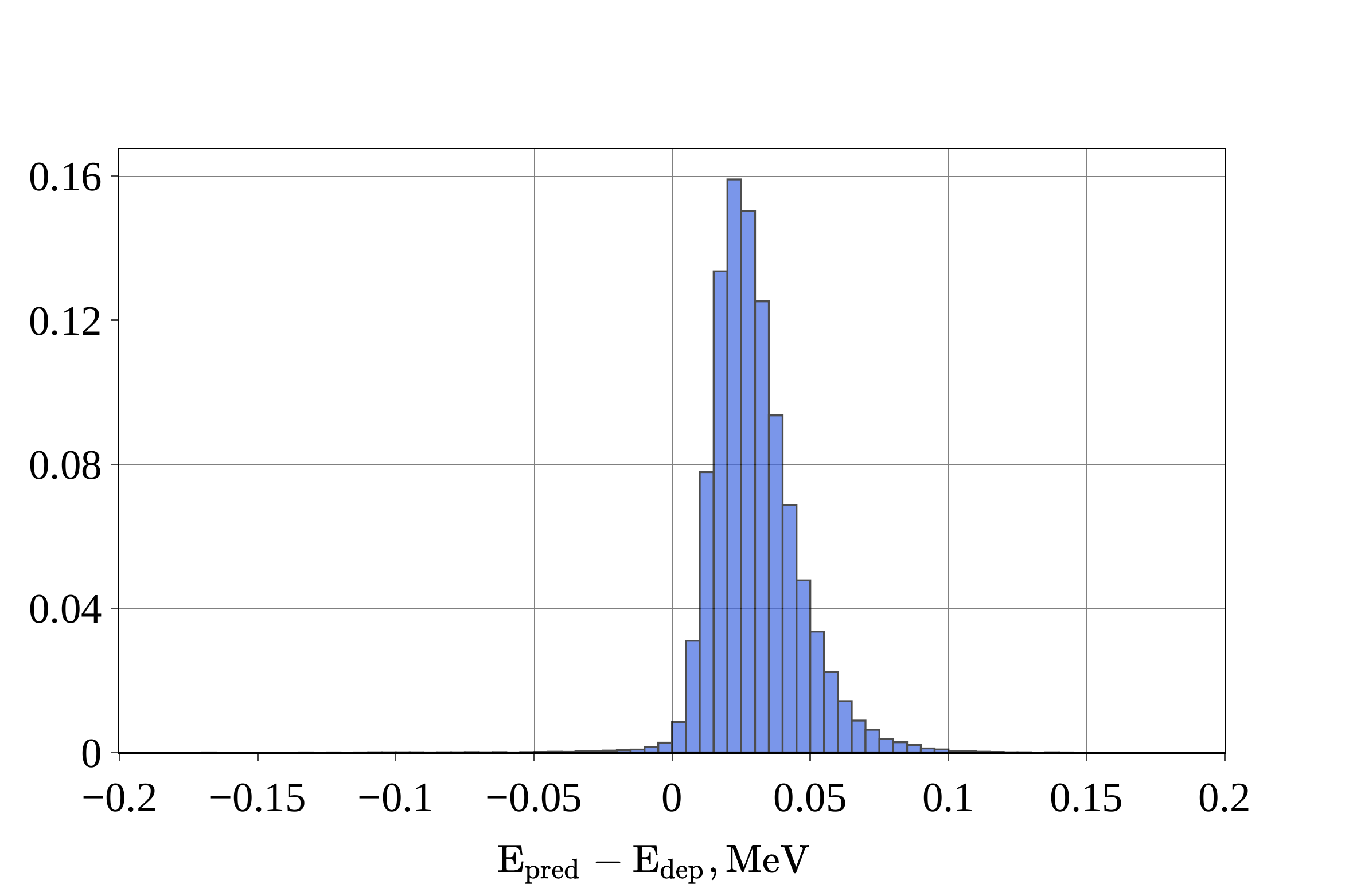}\label{fig:preds_distribution_BDT_1_022_MeV}}
	\caption{Examples of distributions of the predicted deposited energy for ${E_{\rm dep}} = 3.022$~MeV (a) and ${E_{\rm dep}} = 1.022$~MeV (b). The orange line in the left figure represents the Gaussian fit. The distributions were obtained using the BDT model}
	\label{fig:preds_distribution_BDT}
\end{figure*}

As mentioned in Section~\ref{sec:problem}, we use resolution and bias metrics to evaluate the performance of the models. These metrics are obtained from the Gaussian fit of the $E_{\rm pred} - E_{\rm dep}$ distribution as its standard deviation ($\sigma$) and mean value ($\mu$). Figure~\ref{fig:preds_distribution_BDT_3_022_MeV} illustrates an example of the $E_{\rm pred} - E_{\rm dep}$ distribution and its Gaussian fit with $E_{\rm dep} = 3.022$ MeV.
Note that we exclude edge points with the \num{0} MeV and \num{10} MeV kinetic energy corresponding to \num{1.022} and \num{11.022} MeV of deposited energy. ML models learn that the energy of an event belongs to the energy range from the training dataset ($E_{\rm kin} \in 0-10$ MeV) and, therefore, the distributions at the edge points are truncated and exhibit an artificially increased resolution as shown in Figure~\ref{fig:preds_distribution_BDT_1_022_MeV}.

\begin{figure}[!htb]
	\centering
	\includegraphics[width=1\columnwidth]{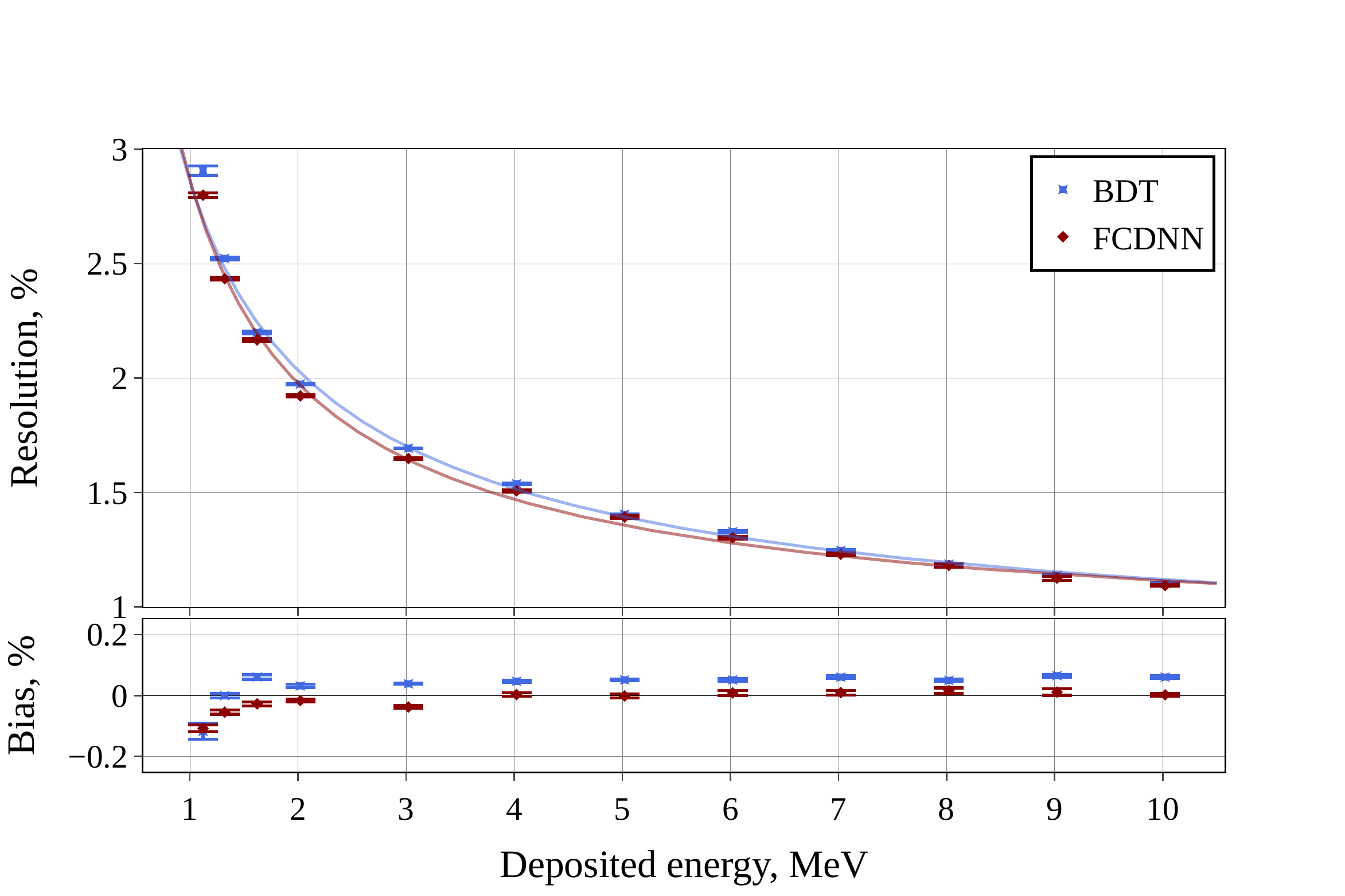}
	\caption{Energy resolution performance: resolution (upper panel) and bias (lower panel) for BDT and FCDNN models. Note that the most left point corresponds to \num{1.122}~MeV}
	\label{fig:BDT_FCDNN_performance}
\end{figure}

Figure~\ref{fig:BDT_FCDNN_performance} shows the energy reconstruction performance (resolution and bias) of the BDT and FCDNN models described in Sections~\ref{sec:mlBDT} and \ref{sec:mlFCDNN}.
Parameterization~\ref{eq:appr_func} with fitted $a$, $b$, and $c$ values is also shown. The energy reconstruction bias for both models is very close to zero.
The resolution performance for FCDNN is better in the lower energy region and converges with the BDT results at higher energies.

Table~\ref{tab:res_params} lists parameters $a$, $b$, and $c$ of Parameterization~\ref{eq:appr_func} of the energy resolution curves and the effective resolution $\tilde{a}$ for both BDT and FCDNN models. The value of $\tilde{a}$ is noticeably better for FCDNN, but for both models $\tilde{a}$ satisfies the requirement of JUNO: $\tilde{a} < 3\%$.

\begin{table}[!htb]
	\centering
	\begin{tabular}{|l|l|l|}
	\hline
    \backslashbox{\it{Parameter}}{\it{Model}}
	& BDT & FCDNN \\
	\hline
	$a \pm \Delta a$ &  2.573 $\pm$ 0.097 & 2.316 $\pm$ 0.139 \\
	\hline
	$b \pm \Delta b$ & 0.763 $\pm$ 0.045 & 0.827 $\pm$ 0.054 \\
	\hline
	$c \pm \Delta c$ & 0.990 $\pm$ 0.394 & 1.474 $\pm$ 0.285 \\
	\hline
	$\tilde{a} \pm \Delta \tilde{a}$ & 2.914 $\pm$ 0.016 & 2.822 $\pm$ 0.027 \\
	\hline
	\end{tabular}
	\caption{Parameters $a$, $b$, and $c$ of parameterization for the energy resolution curves and effective resolution $\tilde{a}$ for BDT and FCDNN models}
	\label{tab:res_params}
\end{table}

\begin{figure}[!htb]
	\centering
	\includegraphics[width=1\columnwidth]{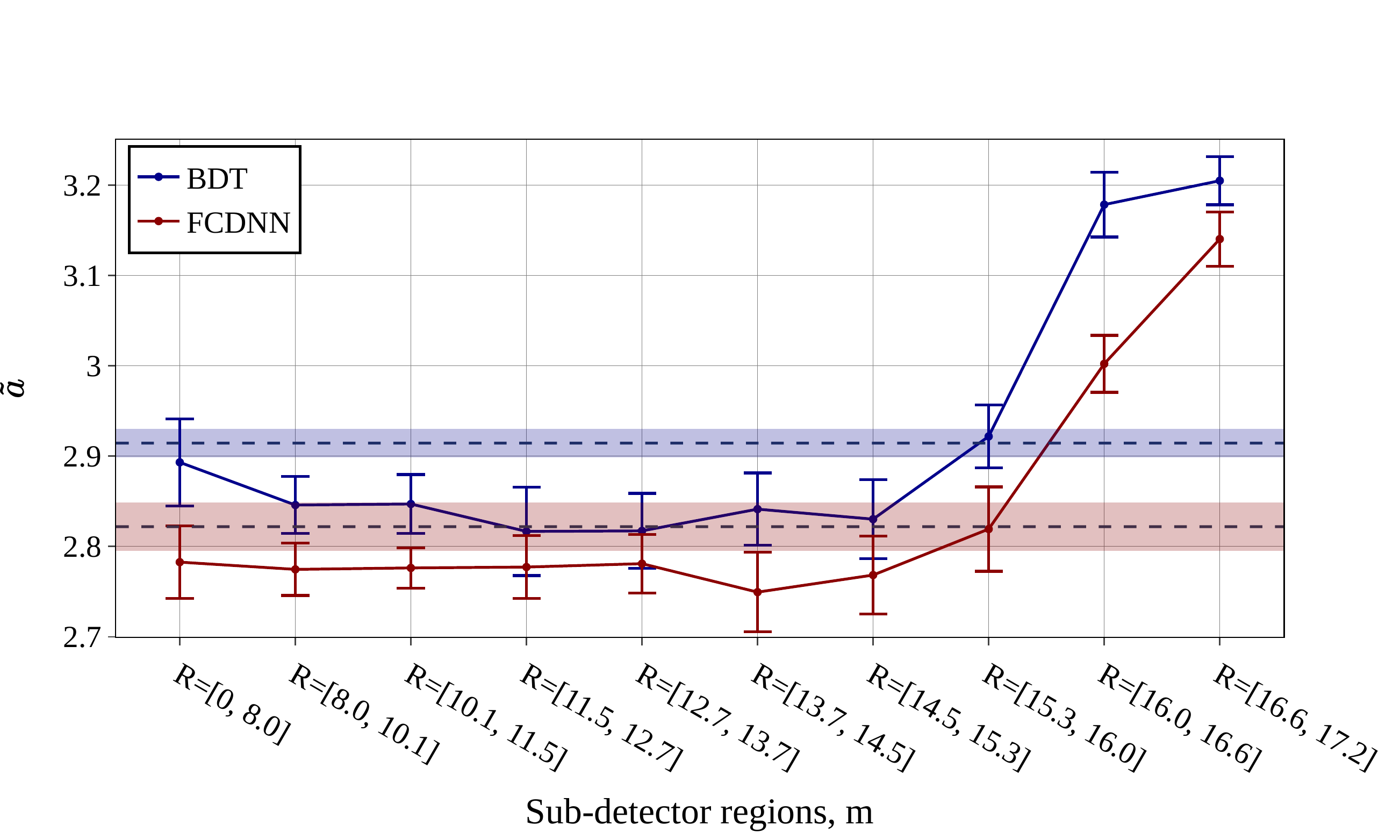}
	\caption{Expected value of the effective resolution $\tilde{a}$ for equidistant subdetector regions. Dashed lines and filled areas represent the effective resolution $\tilde{a}$ and its standard deviation for the BDT and FCDNN models, respectively. Both models are trained on data with $R_{\rm FV} <$ 17.2}
	\label{fig:atilde_vs_regions}
\end{figure}

We also investigate how the resolution depends on different subdetector regions, see Figure~\ref{fig:atilde_vs_regions}. We used our models, BDT and FCDNN, trained on a dataset with the standard fiducial volume cut of 17.2~m, i.e. in the 0--17.2~m range, but tested them on datasets with varied $R$ ranges. The $R$ ranges are selected to contain approximately the same number of events. The effective resolution in the $R$ regions is correlated with the accumulated charge per MeV (compare to Figure~\ref{fig:mean_charge_vs_R}): less charge means a smaller $\tilde{a}$, and vice versa. 
Consequently, the performance of both models worsens for the events close to the edge of the detector.

Besides that, the outer region of the detector is more populated by background events originating from radioactive decays in materials. Therefore, one may consider reducing the fiducial volume to increase the quality of data. On the other hand, excluding outer events decreases statistics, which limits the experiment sensitivity to the neutrino oscillation pattern. Thus, the optimal strategy has to be found via a comprehensive sensitivity study.

\section{Conclusions}
\label{sec:conclusions}

In this work, we present an application of machine learning techniques for precise energy reconstruction in the energy range of 0--10~MeV. We use two models: Boosted Decision Trees (BDT) and Fully Connected Deep Neural Network (FCDNN), trained using aggregated features extracted from Monte Carlo simulation data. 
We considered the case of the JUNO detector. However, the approaches are valid for other similar detectors with a large liquid scintillator target surrounded by an array of photo-sensors. Our dataset is generated with the official JUNO software for modeling particle interactions, light emission, transport, and collection, as well as realistic response of PMTs and the readout system.

We design a large set of features and select one feature subset for the BDT model and another feature subset for the FCDNN model. Both provide the same performances as the full set of features. The requirement on the effective resolution to determine the neutrino mass ordering, $\tilde{a} \leq 3\%$, is achieved by both BDT and FCDNN. BDT is a fast and minimalistic model, making predictions 3-4 times faster than FCDNN. On the other hand, the latter provides a slightly better performance. This follows the trend observed in~\cite{Borisov:2021} where authors found that neural networks trained on large and detailed datasets outperform models based on decision trees. However, for smaller datasets, decision trees have an advantage. For the experiments like JUNO, one can generate enough large datasets, but it is also mandatory to use sparse calibration data. This situation makes the choice between the models not evident, and both approaches should be considered.

\begin{acknowledgement}
We are very thankful to the JUNO collaborators who contributed to the development and validation of the JUNO simulation software. 
We thank N.~Kutovskiy and N.~Balashov for providing an extensive IT support with computing resources of JINR cloud services~\cite{Baranov:2016gvt}, and X.~Zhang for her work on producing MC samples. We are also grateful to Maxim Gonchar for fruitful discussions.
Fedor Ratnikov is supported by the grant for research centers in the field of AI provided by the Analytical Center for the Government of the Russian Federation (ACRF) in accordance with the agreement on the provision of subsidies (identifier of the agreement 000000D730321P5Q0002) and the agreement with the HSE University \textnumero{70-2021-00139}.
Yury Malyshkin is supported by the Russian Science Foundation under grant agreement \textnumero{21-42-00023}.
\end{acknowledgement}

\end{document}